\documentclass[a4paper, 11pt]{article}
\usepackage{amsmath} \usepackage{amsfonts} \usepackage{amssymb}\usepackage{latexsym}
\usepackage[english]{babel}
\usepackage{amscd}
\usepackage[dvips,final]{graphics}
\usepackage{fancyhdr}
\usepackage{locengli,math}
\usepackage{graphicx,pst-plot,psfig,pstricks,pst-node,pst-text,pst-tree,pst-char,pst-coil}

\newcommand{\ind}{\textrm{ind}}
\newcommand{\Ind}{\textrm{Ind}}
\newcommand{\Comp}{\textrm{Comp}}
\newcommand{\PO}{\mathbb{P} \mathbb{O}}
\newcommand{\F}{\mathcal{F}}
\newcommand{\E}{\mathcal{E}}
\begin{document}

\begin{center}
\textbf{\LARGE{\textsf{Coassociative grammar, periodic orbits and quantum random walk over $\mathbb{Z}$}}}
\footnote{
\textit{2000 Mathematics Subject Classification: 16W30; 05C20; 05C90; 60J99; 81Sxx.}
\textit{Key words and phrases:} Coassociative coalgebra, Markov $L$-coalgebra, quantum graphs, De Bruijn graphs, periodic orbits,
chaotic map, random walks, Hadamard walk.}

\vskip1cm
\parbox[t]{14cm}{\large{
Philippe {\sc Leroux}}\\
\vskip4mm
{\footnotesize
\baselineskip=5mm
Institut de Recherche
Math\'ematique, Universit\'e de Rennes I and CNRS UMR 6625\\
Campus de Beaulieu, 35042 Rennes Cedex, France, philippe.leroux@univ-rennes1.fr}}
\end{center}

\vskip1cm
{\small
\vskip1cm
\baselineskip=5mm
\noindent
{\bf Abstract:}
This work will be devoted to the quantisation of
the classical Bernoulli random walk over $\mathbb{Z}$.
As this random walk  is isomorphic to the classical chaotic dynamical system $x \mapsto 2x  \mod 1$ with $x \in [0,1]$,
we will explore the r\^ole
of classical periodic orbits of this chaotic map in relation with a non commutative algebra
associated with the quantisation of the Bernoulli walk.
In particular we show that the set of periodic
orbits, $\PO$, of the map $x \mapsto 2x  \mod 1$ can be embeded into a language equipped with a coassociative grammar and
for any fixed time,
that any vertex of $\mathbb{Z}$ is in one to one with of a subset of $\PO$.  The reading and the contraction maps applied to
these periodic orbits
allow us to recover the combinatorics generated by the quantum random walk over  $\mathbb{Z}$
\section{Introduction and Notation}
Motivated by the success of classical random walks and chaotic dynamical systems, we study the quantisation of the
random walk over $\mathbb{Z}$ and its relationships with a classical chaotic system $x \mapsto 2x  \mod 1$ with $x \in [0,1]$ .
The first part will be devoted to notation and to the introduction of notions required in the sequel. We recall the notion of
quantum graphs, unistochastic processes, directed graphs and extension, $L$-coalgebras and coproducts, the relation between
the Bernoulli walk and the map $x \mapsto 2x \mod 1$ and a quantisation of this walk proposed by Biane \cite{Biane}.
Section 2 displays the relationships between some coassociative coalgebras, chaotic maps $x \mapsto nx \mod 1$, with $n>1$
and $(n,1)$-De Bruijn graphs. Section 3 yields a bridge between quantum channels and quantum graphs. Section 4 defines
the quantum random walk over $\mathbb{Z}$ and section 5 explains the link between periodic orbits of $x \mapsto 2x \mod 1$,
their coassociative language and the combinatorics generated by the quantum random walk over $\mathbb{Z}$.
\subsection{Quantum graphs}
\label{assump}
Let $B$ be a bistochastic matrix representing a directed graph, i.e. two vertices $x_i$ and $x_j$ are linked iff $B_{ij} \not= 0$.
$B$ is said unistochastic if it exists an unitary matrix $U$ such that $B_{ij}=\vert U_{ij} \vert^2$. In this case,
we say that the graph can be quantized. The notion of quantum graphs were introduced as a toy model for studying quantum chaos by Kottos
and Smilansky \cite{Kottos}, \cite{Kottos2}. This notion was also studied by Tanner \cite{Tanner} and by Barra and Gaspard \cite{BG} \cite{BG2}. In this article we will follow
another appraoch leading to quantum graphs puts forward in \cite{Kus} concerning the one dimensional dynamical systems. They consider
an one-dimensional mapping $f$ acting on $I=[0,1]$ such that $f: I \xrightarrow{} I$, is piecewise linear.
Moreover $f$ verifies the following three conditions:
\begin{enumerate}
\label{cond}
\item {There exists a Markov partition of the interval $I$ into $M$ equal cells $E_i :=[\frac{i-1}{M}, \frac{i}{M}), \ i=1 \ldots M$, with $M$
a positive integer and $f$ is linear on each cell $E_i$.}
\item{For all $y \in I$, $$\sum_{x \in f^{-1}({y})} \frac{1}{f'(x)}=1,$$
where $f'$ is the right derivative, (defined almost everywhere on $I$).}
\item{The finite transfer matrix $B$, describing the action of $f$ on the cells $E_i$ is unistochastic.}
\end{enumerate}

\Rk
On each cell, $f$ coïncides with the function $f_i: [\frac{i-1}{M}, \frac{i}{M}) \xrightarrow{} I$, $x \mapsto c_ix +b_i$ where the $c_i$ have to be
 non zero integers and the $b_i$ are rational. The unistochastic matrix $B$ is a $M$ by $M$ matrix and $B_{ij} = \frac{1}{\vert c_i \vert}$.
 Thus the probability of visiting the cell $E_j$ from $E_i$ is equal to $\frac{1}{f'(x)}$, with $x \in E_i$ and $f(x) \in E_j$.

\Rk
The Kolmogorov-Sinai-entropy of the Markov chain generated by the bistochatsic matrix $B$ \cite{Katok} is
$$ H_{\textrm{KS}}=-\sum_{i=1}^M \tilde{p_i} \sum_{j=1}^M B_{ij} \  \textrm{log} \ B_{ij},$$
where $\tilde{p}$ is the normalized left eigenvector of $B$ such that $\tilde{p}B = \tilde{p}$, with $\sum_{i=1}^M \tilde{p}_i =1.$
This equation gives the dynamical entropy of the system since the Markov partition on $M$ equal cells is a generating partition of the system.
As the transition matrix is bistochastic, all the components of $\tilde{p}$ are equal to $\frac{1}{M}$. Thus $H_{\textrm{KS}}=0$ iff
all the $B_{ij} \in \{ 0,1 \}$. This entails that $\vert f'(x) \vert=1$, i.e. the system is regular. With the conditions stated above, the converse is true.

\begin{exam}{[Regular system \cite{Kus}]}\\
Here is an example of such piecewise linear map. The associated bistoschastic matrix is
$ B_3 =\begin{pmatrix}
0 & 0 & 1 \\
1 & 0 & 0 \\
0 & 1 & 0 \\
\end{pmatrix}$.
\begin{center}
\includegraphics*[width=6cm]{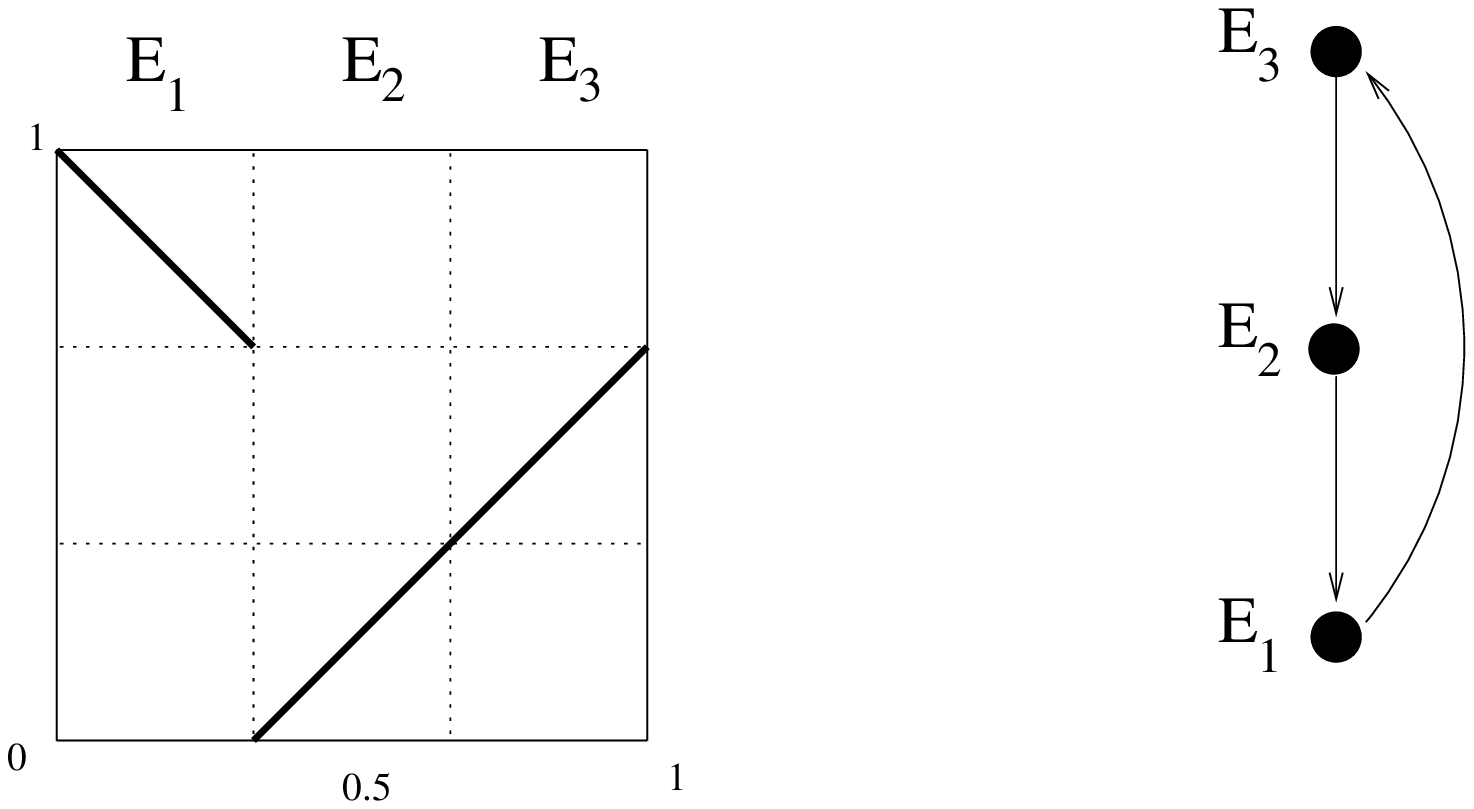}
\end{center}
\end{exam}
In the sequel we will be interested by one of the simplest 1D maps displaying chaotic dynamics.
\begin{exam}{[Chaotic system, the Bernoulli shift]}\\
The Bernouilli shift is described by $f: x \mapsto 2x  \mod 1$. This map is associated with the unistochastic matrix $B_2 = \frac{1}{2}\begin{pmatrix}
1 & 1 \\
1 & 1 \\
\end{pmatrix}$.
\begin{center}
\includegraphics*[width=6cm]{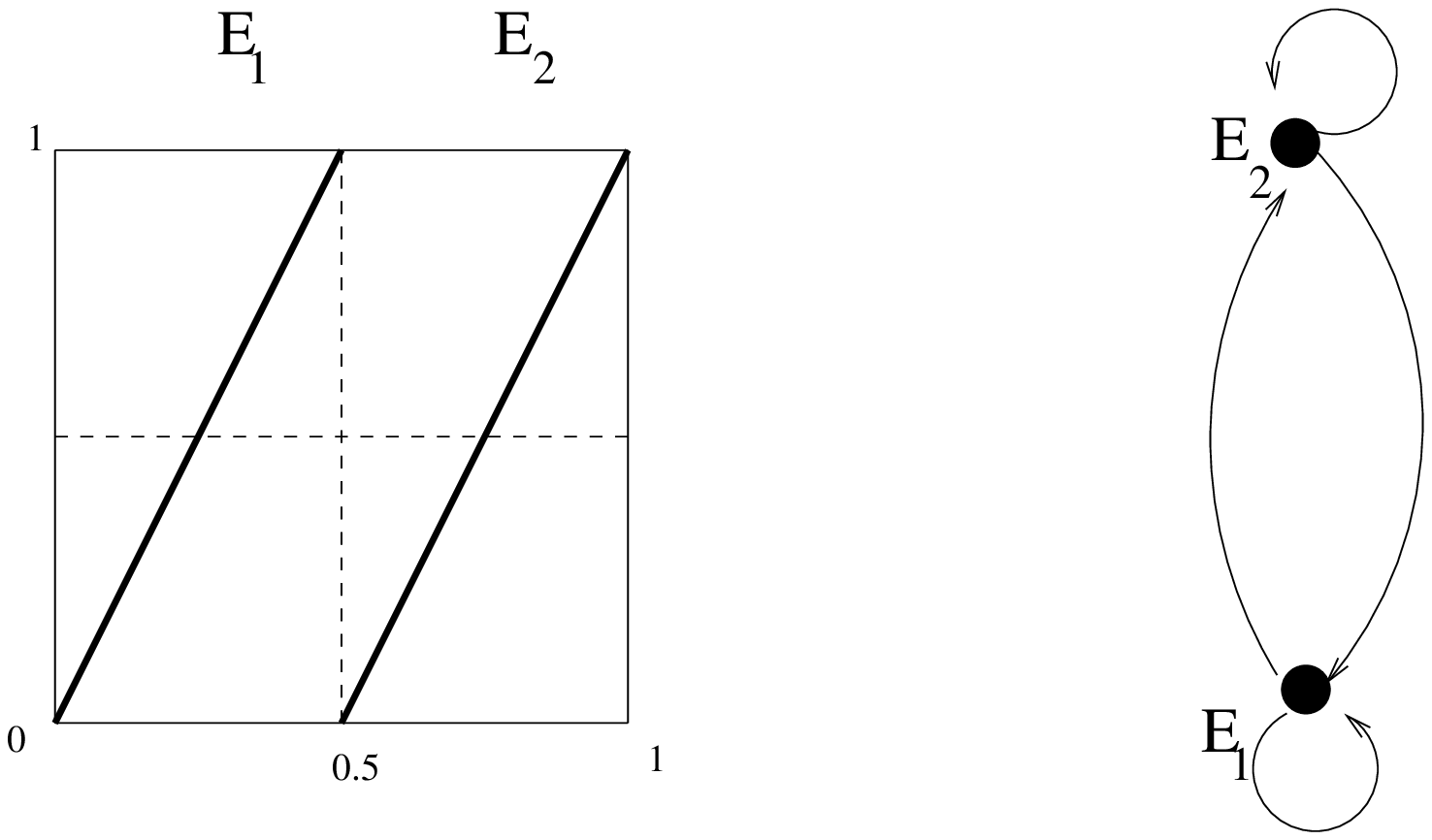}
\end{center}
\end{exam}
We recall an important theorem in \cite{Kus}.
\begin{theo}
With the assumptions described in \ref{assump} on the piecewise linear map $f$,  to every
periodic orbit of period $n$ of the dynamical system described by $f$,
corresponds an unique periodic orbit of period $n$ of the directed graph describing by the associated unistochastic matrix.
\end{theo}
\subsection{$L$-coalgebras}
To unify the directed graphs framework, even equipped with a family of probability vectors,  with coassociative coalgebra theory,
we are led to
introduce the notion of $L$-coalgebra over a field $k$ \cite{Coa}, i.e. a $k$-vector space equipped with two coproducts
 $\Delta$ and $\tilde{\Delta}$ which obey
the coassociativity breaking equation
$(\tilde{\Delta} \otimes id)\Delta = (id \otimes \Delta)\tilde{\Delta}$.
If $\Delta = \tilde{\Delta}$, the $L$-coalgebra is said degenerate. Moreover a $L$-coalgebra can have two counits, the right
counit $\epsilon: G \xrightarrow{} k$ which verifies $ (id \otimes \epsilon)\Delta = id$
and the left counit $\tilde{\epsilon}: G \xrightarrow{} k$, which verifies $ ( \tilde{\epsilon} \otimes id)\tilde{\Delta} = id. $

One of the interests of such a formalism is to describe directed graphs \footnote{Here supposed to be with no sources and no sinks, see \cite{Coa} otherwise.}, equipped
with a family of probability vectors or not, thanks to their coproducts instead of the classical source and terminus mappings.
For building such a directed graph for each $L$-coalgebra, we associate with each tensor product $x \otimes y$ appearing in the definition of
the coproducts a directed arrow $x \xrightarrow{}y$. For instance, here is the directed graph associated with $Sl_q(2)$.
\begin{exam}{[$Sl_q(2)$]}
The well-known coassociative coalgebra structure is:
$$
\Delta a = a \otimes a + b \otimes c, \ \
\Delta b = a \otimes b + b \otimes d, \ \
\Delta c = d \otimes c + c \otimes a, \ \
\Delta d = d \otimes d + c \otimes b.
$$
and its directed graph is  $G(Sl_q(2))$:
\begin{center}
\includegraphics*[width=4cm]{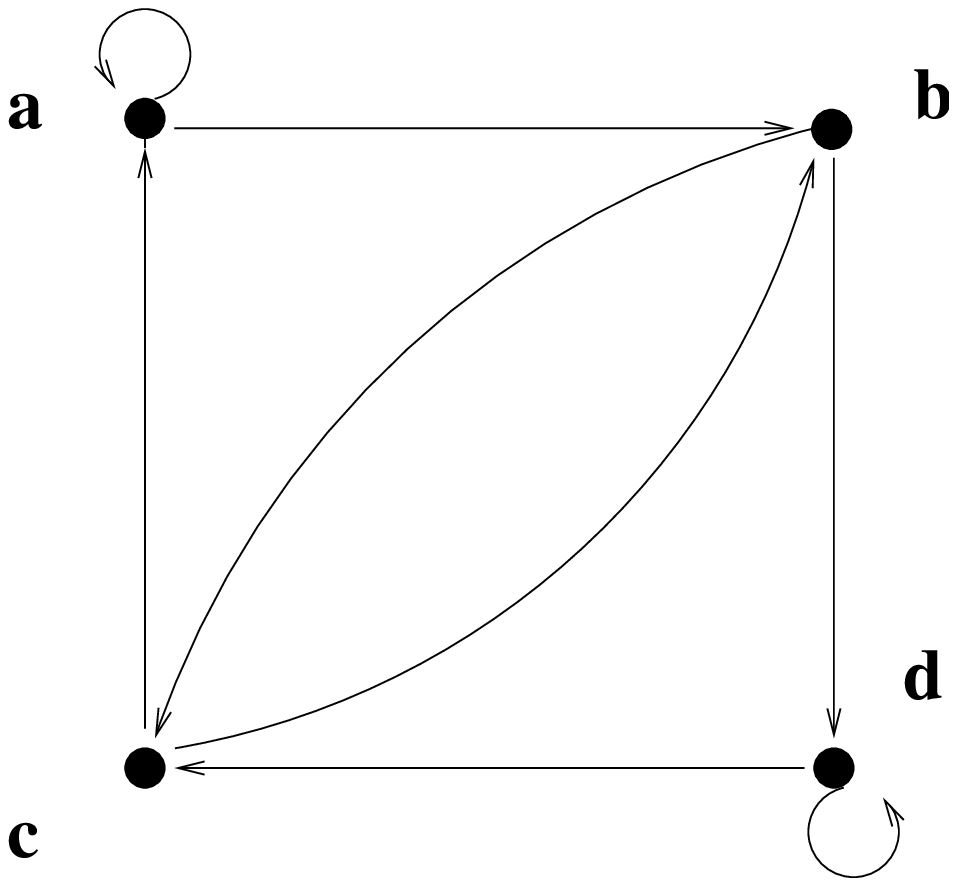}
\end{center}
\end{exam}
\Rk
In the sequel, we denote by $\E$ the coassociative coalgebra generated by $a,b,c$ and $d$, having the same coproduct and counit
definitions as $Sl_q(2)$.
\begin{defi}{[Markov $L$-coalgebra]}
A {\it{Markov $L$-coalgebra}} $C$ is a $L$-coalgebra such that for all $v \in C $, $\Delta v = \sum_{i \in I} \lambda_i v \otimes v_i $
and $ \tilde{\Delta}v =\sum_{j \in J} \mu_j v_j \otimes v $, with $v_i,v_j \in C$, $\lambda_i, \mu_j \in k$ and $I,J$ are finite sets.
\end{defi}
Such a structure reproduces locally what we have in mind when we speak about random walks on
a directed graph if the scalars are positive and the right counit $v \mapsto 1$ exists.
We recall that the right counit is a map $\epsilon: C \xrightarrow{} k$ verifying $(id \otimes \epsilon) \Delta = id$.
A {\it{Markov $L$-bialgebra}} is a Markov $L$-coalgebra and an unital algebra such that its coproducts and counits are homomorphisms.
\begin{exam}{[The graph of the map $x \mapsto 2x \mod 1$]}
\begin{center}
\includegraphics*[width=4cm]{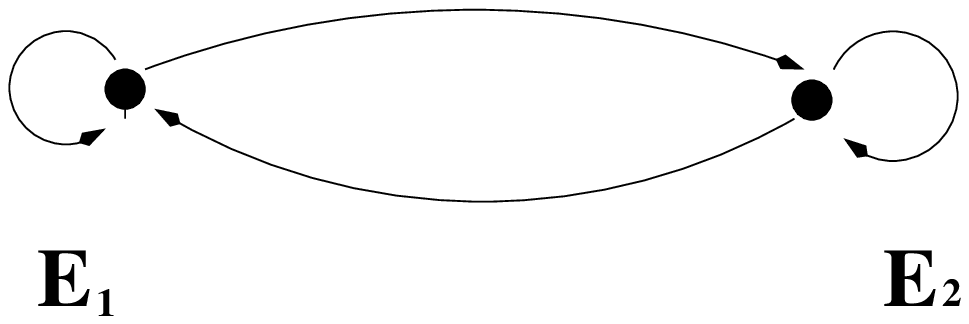}
\end{center}
With the $(2,1)$-De Bruijn graph, we associate its natural Markov $L$-coalgebra by defining:
$\Delta E_1 = E_1 \otimes E_1 + E_1 \otimes E_2$ and $\Delta E_2 = E_2 \otimes E_1 + E_2 \otimes E_2$,
$\tilde{\Delta}E_1 =E_2 \otimes E_1 + E_1 \otimes E_1$ and $\tilde{\Delta}E_2 =E_1 \otimes E_2 + E_2 \otimes E_2$.
\end{exam}
\begin{exam}{[Unital algebra]}
Let $A$ be an unital algebra. $A$ carries a non trivial Markov $L$-bialgebra called the flower graph
with coproducts $\delta(a) = a \otimes 1$ and $\tilde{\delta}(a)= 1 \otimes a$, for all $a \in A$.
We call such a Markov $L$-coalgebra a \textit{flower graph} because it is the concatenation of petals:
\begin{center}
\includegraphics*[width=5cm]{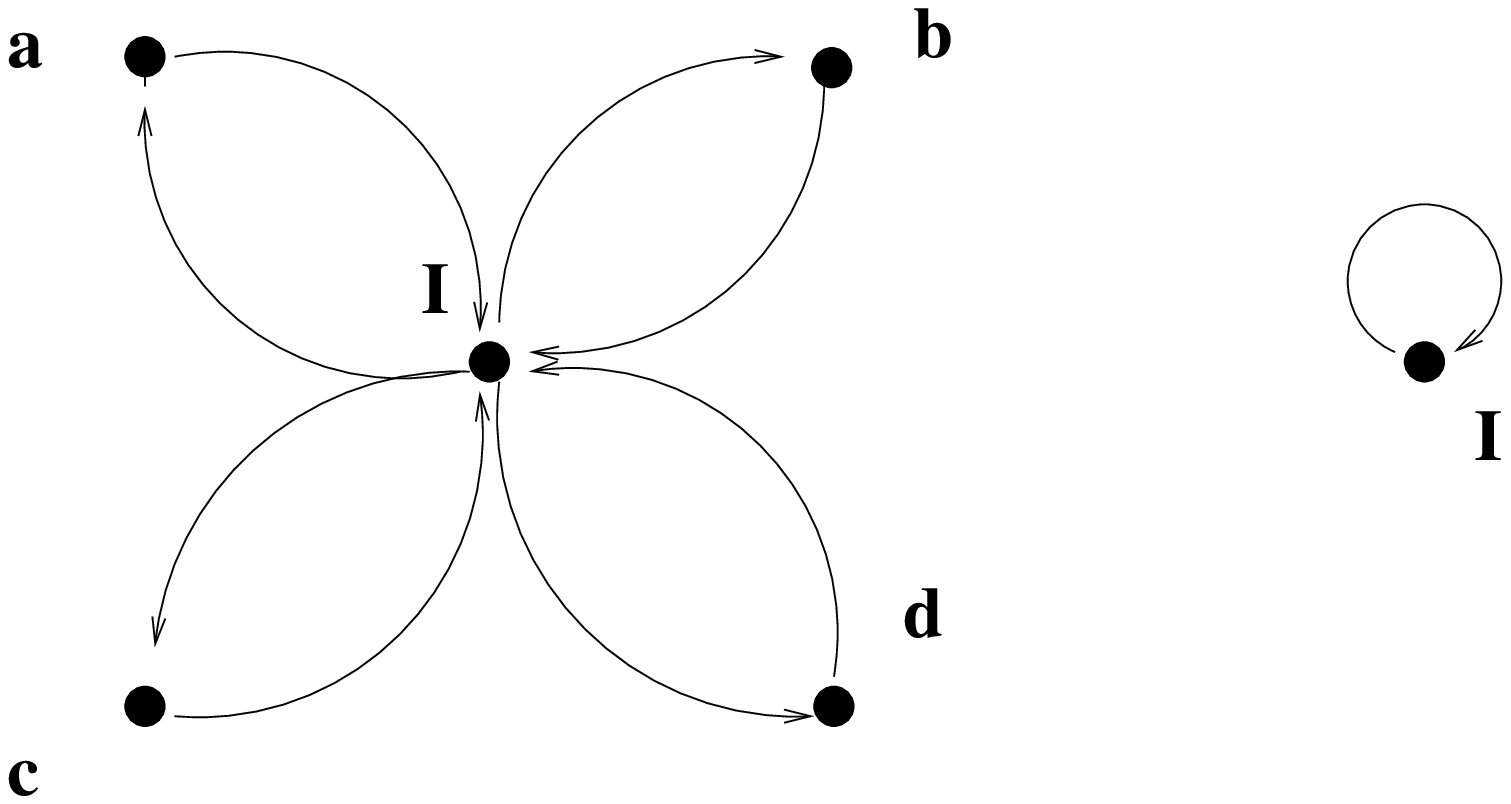}
\end{center}
\end{exam}
An other example is the directed triangle graph:
\begin{center}
\includegraphics*[width=5cm]{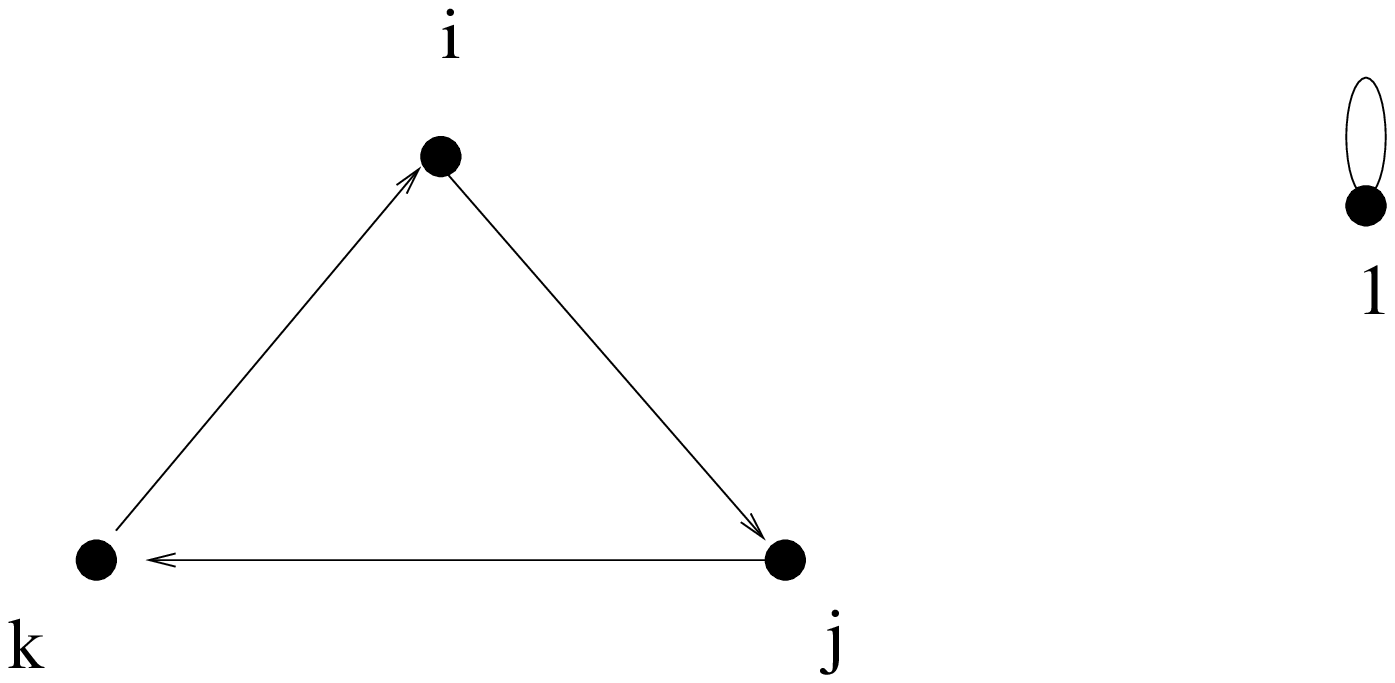}
\end{center}
Defining $x_0 \equiv i$, $x_1 \equiv j$, $x_2 \equiv k$ and adding subscripts
$\alpha, \beta \in \{0,1,2 \} \ \ \textrm{mod} \ 3$ i.e. $x_{\alpha+ \beta}
\equiv x_{(\alpha+ \beta) \textrm{mod} \ 3}$, we define,
$\Delta  x_{\alpha} = x_{\alpha} \otimes x_{\alpha +1}, \ \ \Delta 1 = \tilde{\Delta} 1 = 1 \otimes 1, \ \ \tilde{\Delta}x_{\alpha} = x_{\alpha -1} \otimes x_{\alpha}$.
They embed the directed triangle graph into a Markov $L$-coalgebra with counits. It is used to embed the quaternions algebra or $M_2(k)$, i.e.
the algebra generated by the Pauli matrices into a Markov $L$-Hopf algebra (of degree 2).

We recall that the sequence,
$ \Delta_1 \equiv \Delta, \Delta_2 = id \otimes \Delta, \Delta_3 = id \otimes id \otimes \Delta, \ldots $,
generates all possible (random) walks starting at any vertex.
Similarly, the sequence of powers of
$\tilde{\Delta}$, generates all the possible (random) walks arriving at a given vertex.

\subsection{Classical random walk over $\mathbb{Z}$ and the Bernoulli shift}
We consider the random walk over $\mathbb{Z}$, i.e. we consider
$ \Omega= \{-1,1 \}^{\mathbb{N}}$ equipped with the product measure $\mu^{\otimes \mathbb{N}}$, where $\mu = \frac{1}{2}\delta_{-1} +
\frac{1}{2}\delta_1$. We consider the sequence of iid random variables $(X_n)_{n \in \mathbb{N}}$,
with:
$$ X_n: \Omega \xrightarrow{} \{ -1, +1 \}, \ \ \textrm{such that} \  X_n(\omega) = \omega_n.$$
It is well known that this process and the
symbolic dynamics generated by $x \mapsto 2x \mod 1$ are isomorphic. Indeed consider the iid process defined by
$ Y_n: \Omega \xrightarrow{} \{ 0, +1 \}, \ \ \textrm{such that} \  Y_n(\omega) = \omega_n' := \frac{\omega_n +1}{2}$.
$ \Omega= \{-1,1 \}^{\mathbb{N}}$ becomes $ \Omega'= \{0,1 \}^{\mathbb{N}}$ and $\mu$ becomes
$\mu' = \frac{1}{2}\delta_{0} +
\frac{1}{2}\delta_1$.
Define the mesurable function:
$$ \Phi: \Omega' \xrightarrow{} [0,1[, \ \ \ \omega' \mapsto \sum_{n=0}^\infty \ \frac{\omega_n' }{2^{n+1}}.$$

The cylinder $C = [Y_0 = \omega_0' ; \ldots; Y_l=\omega_l']$, being the set of the sequences starting by $(\omega_0'; \ldots; \omega_l')$,
will cover the interval $[\sum_{n=0}^l \ \frac{\omega_n'}{2^{n+1}}; \sum_{n=0}^l \ \frac{\omega_n'}{2^{n+1}} +
\sum_{n=l+1}^\infty \ \frac{1}{2^{n+1}}]$. We notice that Leb($C$) $=\frac{1}{2^{l+1}}$, where Leb is the Lebesgue measure.
Let us consider the shift $\theta$ such that $(\theta \omega')_n = \omega_{n+1}'$. This shift lets the Lebesgue measure of the cylinder $C$
invariant if we denote $\theta( [Y_0 = \omega_0'; \ldots; Y_l=\omega_l'])=( [Y_1 = \omega_0'; \ldots; Y_{l+1}=\omega_l'])$.
Moreover we have $\Phi( \theta(\omega')) = 2 \Phi (\omega') \mod 1$.

The random walk over $\mathbb{Z}$, described by
($\Omega= \{-1,1 \} ^\mathbb{N}, \mu^{\otimes \mathbb{N} }, \theta$) is isomorphic to
($\Omega'= \{0,1 \} ^\mathbb{N}, \mu'^{\otimes \mathbb{N} }, \theta'$)
which is isomorphic to the chaotic system ($[0,1[, \beta [0,1[, \textrm{Leb}, f: x \mapsto 2x \mod 1$). As $\Phi \circ \theta' = f \circ \Phi$,
the following diagram,
\begin{equation*}
\begin{CD}
 (\{0,1 \}^{\mathbb{N}}, \mu^{\otimes \mathbb{N} })@>\theta>> (\{0,1 \}^{\mathbb{N}}, \mu^{\otimes \mathbb{N} }) \\
@V{\Phi}VV		@VV{\Phi}V \\
([0,1[, \textrm{Leb}) @>{f }>> ([0,1[, \textrm{Leb})
\end{CD}
\end{equation*}
is commutative.
\subsection{quantisation of the classical Bernoulli walk}
In \cite{Biane}, Biane purposes a non commutative version of the Bernoulli process. Set $\Omega:= \{ +1, \ -1 \}$, the probability space
and define the probability
$\mathbb{P}( \{ +1 \}) = p$ and $\mathbb{P}( \{ -1 \}) = q$.
The process $X: \Omega \xrightarrow{} \mathbb{R}$ is defined as $X(+1) = +1$ and $X(-1) = -1$.
By identifying $(1,0)$ with $(4p)^{-\frac{1}{2}}(1+X)$ and $(0,1)$ with $(4q)^{-\frac{1}{2}}(1-X)$,
the space $L^2(\Omega, \mathbb{P})$ is isomorphic to $\mathbb{C}^2$. We notice that the algebra $L^\infty (\Omega, \mathbb{P})$,
acting on $L^2$ can be identified with diagonal matrix of $M_2(\mathbb{C})$. A natural non commutative generalisation consists in
lifting this commutative algebra into a bigger non commutative one, i.e. $M_2(\mathbb{C})$.
In the sequel, we will not follow the Biane's work. Nevertheless, we keep in mind that $M_2(\mathbb{C})$ is a suitable algebra and
we will see later how to rediscover it.
\section{Relationships between coassociative coalgebras and chaotic maps $x \mapsto nx \mod 1$}
\begin{defi}{[De Bruijn graph]}
A $(p,n)$ De Bruijn sequence on the alphabet $\Sigma = \{ a_1, \ldots, a_p \}$ is a sequence $(s_1, \ldots, s_{m})$ of $m = p^n$ elements
$s_i \in \Sigma$ such that subsequences of length $n$ of the form $(s_i, \ldots, s_{i+n -1})$  are distinct,
the addition of subscripts being done modulo $m$.
A {\it{$(p,n)$-De Bruijn graph}} is a directed graph
whose vertices correspond to all possible strings $s_1s_2 \ldots s_n$ of $n$ symbols from $\Sigma$.
There are $p$ arcs leaving the vertex $s_1s_2 \ldots s_n$ and leading to the adjacent node $s_2s_3 \ldots s_n \alpha$, $\alpha \in \Sigma$.
Therefore the {\it{$(p,1)$-De Bruijn graph}} is the directed graph with $p$ vertices, complete, with a loop at each vertex.
\end{defi}
\begin{defi}{[Extension]}
The {\it{extension}} of a directed graph $G$, with vertex set $J_0= \{j_1, \ldots, j_n \}$ and edges set $A_0 \subseteq J_0 \times J_0$
is the directed graph with vertex set $J_1 = A_0$ and the edge set $A_1 \subseteq J_1 \times J_1$ defined by
$((j_k, j_l), (j_e, j_f))$ iff $j_l = j_e$. This directed graph, called the line directed graph, is denoted $E(G)$.
\end{defi}
Recall that the definition of an associative dialgebra is a notion due to Loday \cite{Loday}. Here, we are interested in the
notion of a coassociative co-dialgebra.
\begin{defi}{[Coassociative co-dialgebra of degree $n$]}
Let $D$ be $K$-vector space, where $K$ denotes the real or complex field. For every $n >0$,
Let $\Delta$ and $\tilde{\Delta}$ be two linear mapping $D^{\otimes n} \xrightarrow{}D^{\otimes n+1}$.
$D$ is said a {\it{coassociative co-dialgebra}} of degree $n$ if the following axioms are verified:
\begin{enumerate}
\item {$\Delta$ and $\tilde{\Delta}$ are coassociative,}
\item{$(id \otimes \Delta)\Delta = (id \otimes \tilde{\Delta}) \Delta $,}
\item{$(\tilde{\Delta} \otimes id) \tilde{\Delta} = (\Delta  \otimes id) \tilde{\Delta}$,}
\item{$(\tilde{\Delta} \otimes id) \Delta = ( id \otimes \Delta) \tilde{\Delta}$.}
\end{enumerate}
The last equation is called the coassociativity breaking equation in \cite{Coa}.
\end{defi}
\begin{prop}
The $(n,1)$-De Bruijn graph describes the chaotic dynamics $x \mapsto nx \mod 1$.
\end{prop}
\Proof
The map $x \mapsto nx \mod 1$ is coded by the unistochastic $n$ by $n$ matrix $B_n$, with $B_{ij} =\frac{1}{n}$, for all $i,j =1, \ldots n$.
The associated directed graph is then the  $(n,1)$-De Bruijn graph.
\eproof
\begin{prop}
The coproducts of the Markov $L$-coalgebra $G$ of the $(n,1)$-De Bruijn graph define a coassociative co-dialgebra of degree 1.
\end{prop}
\Proof
The coproducts are coassociative and send the Markov $L$-coalgebra $G$ into $G^{\otimes 2}$. The other axioms are straightforward.
There is a right counit $\epsilon$ which maps each vertex into $\frac{1}{n}$ for the coproduct $\Delta$ and a left counit $\tilde{\epsilon}$
which maps each vertex into $\frac{1}{n}$ for the coproduct $\tilde{\Delta}$.
\eproof
\begin{prop}
The extension of the $(n,1)$-De Bruijn graph,
can be equipped with a coassociative coproduct.
\end{prop}
\Proof
Let us denote the edge emerging from a given vertex $i$, with $i = 1, \ldots, n$ of the $(n,1)$-De Bruijn graph $G_n$ by $a_{ij}$.
The new vertex of the extension of $G_n$ is denoted by $a_{ij}$ and the edges are denoted by $((ij), (jk))$. By denoting
$\Delta a_{ij}= \sum_l \ a_{il}\otimes a_{lj}$, this coproduct is coassociative and the graph associating with
the coassociative coalgebra $( \{ a_{ij} \}_{(i,j =1, \ldots n)}, \Delta)$ is easily seen to be $E(G_n)$. It has an obvious counit, $a_{ij} \mapsto 0$ if
$i \not= j$ and $a_{ij} \mapsto 1$ otherwise.
\eproof
\begin{coro}
The extension of the $(2,1)$-De Bruijn graph can be equipped with the coassociative coproduct of
the coassociative coalgebra $\E$.
\end{coro}
\Proof
Straightforward.
\eproof

We have shown that
each Markov $L$-coalgebra described by the $(n,1)$-De Bruijn graph is an unistochastic map associated with the classical chaotic map
$x \mapsto nx \mod 1$. It has a structure of coassociative co-dialgebra. Its extension yields a coassociative coalgebra. The relationships
between the $(n,1)$-De Bruijn graphs and their extensions are treated in more details in \cite{codialg1}.
\section{Quantum channels from quantum graphs}
Let $B$ be a $n$ by $n$ unistochastic matrix associating with a classical dynamical system satisfying the condition \ref{cond}.
We introduce now the quantisation and deformation of the Markov partition associated
with the Markov family $E_i, i=1, \ldots, M$ and the unistochastic
matrix.
The unistochastic matrix $B$ can be decomposed into $n$ independent matrices $X_h$, $h = 1, \ldots n$ defined by $(X_h)_{ij} = B_{ij}$,
if $h=i$ and $0$ otherwise. We get $B = \sum_{h = 1}^k \ X_h$ and
the $X_h$ verify the algebraic relations $X_hX_l = B_{hl}X_l $. When we quantify the directed graphs associated with the
unistochastic matrix $B$, i.e. when we choose an unitary matrix $U$ satisfying $B_{ij}=\vert U_{ij} \vert^2$, we deform the
algebraic relations between the $X_h$ to have $U =\sum_{h = 1}^k \ Q_h$, with $(X_h)_{ij}=\vert (Q_h)_{ij} \vert^2$ and
$Q_hQ_l = U_{hl}Q_l$. We can say that we have quantised the classical dynamics described by the unistochastic matrix $B$.
\begin{exam}{}
The operators $X_h$, $h =1, \ldots k$, associated with the chaotic map $x \mapsto nx \mod 1$ obey the
algebraic laws $X_h^2 =\frac{1}{n^2}X_h$ and $X_hX_lX_h = \frac{1}{n^3}X_h$.
\end{exam}
\begin{prop}
Let $X_h$, $h =1, \ldots k$, be the operators associated with the unistochastic matrix $B$. Its quantisation yields the operators
$Q_h$.
We get $\sum_{h = 1}^k \ Q_hQ_h^{\dagger}=Id$ and $\sum_{h = 1}^k \ Q_h^{\dagger}Q_h=Id$, i.e. the linear map
$ \rho \mapsto \sum_{h = 1}^k \ Q_h^{\dagger} \rho Q_h$ and $ \rho \mapsto \sum_{h = 1}^k \ Q_h \rho Q_h^{\dagger}$ are
quantum channels. Moreover $Q_lQ_h^{\dagger} = 0$ if $h \not= l$, i.e.
the quantum channel $ \rho \mapsto \sum_{h = 1}^k \ Q_h \rho Q_h^{\dagger}$ is an homomorphism.
\end{prop}
\Proof
This is due to the fact that the columns of $U$ are orthonormal.
\eproof

\noindent
In the following we will focus on the $(2,1)$-De Bruijn graph, associated with the decomposition of the
unistochastic map $B_2$ and hence with the chaotic map $x \mapsto 2x \mod 1$.
\Rk
Let $X_1, \ X_2$ be the operators associated with the chaotic map $x \mapsto 2x \mod 1$. One of the possible quantisation of this
chaotic map is the Hadamard matrix,
$ U_{\textrm{H}} = \frac{1}{\sqrt{2}}\begin{pmatrix}
 1 & 1\\
1 & -1
\end{pmatrix}.
$
It is decomposed into two operators,
$ P := Q_1 = \frac{1}{\sqrt{2}}\begin{pmatrix}
 1 & 1\\
0 & 0
\end{pmatrix}$
and $Q := Q_2 = \frac{1}{\sqrt{2}}\begin{pmatrix}
 0 & 0\\
1 & -1
\end{pmatrix}.$
These different choices lead to a set of suitable unitary matrices.
However, for the study of quantum random walk, we can enlarge
this set to include all possible unitary matrices.
In general, such an unitary matrix reads
$U = \frac{1}{\sqrt{2}}\begin{pmatrix}
 \alpha & \beta\\
\gamma &  \delta
\end{pmatrix},$
with obvious conditions on  $\alpha, \beta,
\gamma, \delta$.
Its decomposition is denoted by:
$P = \frac{1}{\sqrt{2}}\begin{pmatrix}
 \alpha & \beta\\
0 & 0
\end{pmatrix}$
and
$Q= \frac{1}{\sqrt{2}}\begin{pmatrix}
 0 & 0\\
\gamma & \delta
\end{pmatrix}$
and verify the following algebraic relations $P^2 = \alpha P, \ Q^2 = \delta Q, \ PQP = \beta \gamma  P, \ QPQ = \beta \gamma Q$.
\begin{prop}
Suppose $\alpha \delta \not=0$. Consider $e_1 := \frac{1}{\alpha}P$ and $e_2 := \frac{1}{\delta}Q$.
We get $e_1^2 =e_1$, $e_2^2 =e_2$, $e_1e_2e_1 = \lambda e_1$ and  $e_2e_1e_2 = \lambda e_2$, where
$\lambda := \frac{\gamma \beta}{\delta\alpha}$, i.e. the algebra generated by $e_1, e_2$ is a Jones algebra \cite{Jones}.
\end{prop}
\Proof
Straightforward.
\eproof
\section{Quantum random walk over $\mathbb{Z}$}
In the physics literature, quantum random walk has been studied for instance, by Ambainis and al \cite{Ambainis},
Konno and al,  in \cite{Konno} \cite{Konno1}. Here we propose a more mathematical
framework for the quantum random walk over $\mathbb{Z}$ and
show that the combinatorics \footnote{We keep the notation of \cite{Konno}.} of this walk can be recovered by using the coproduct of $\E$. We will show that this walk
is closely related to the quantisation of the $(2,1)$-De Bruijn graph and that the polynomials involved in each vertex of $\mathbb{Z}$ are
related, for a given time, in a bijection way,
to periodic orbits of the chaotic map $x \mapsto 2x \mod 1$ and that these periodic orbits can be manipulted by
the coproduct of $\E$.

Let $\mathcal{H}$ be a separable Hilbert space of infinite dimension with $( \ \vert e_n \ket \ )_{n \in \mathbb{Z}}$ as an orthonormal basis.
In \cite{Coa} directed graphs have been put on algebraic structures such as $k$-vector spaces, algebras, coalgebras and so on.
We decide to put on  $\mathcal{H}$, the following directed graph $G_\mathbb{Z}$:
\begin{center}
\includegraphics*[width=10cm]{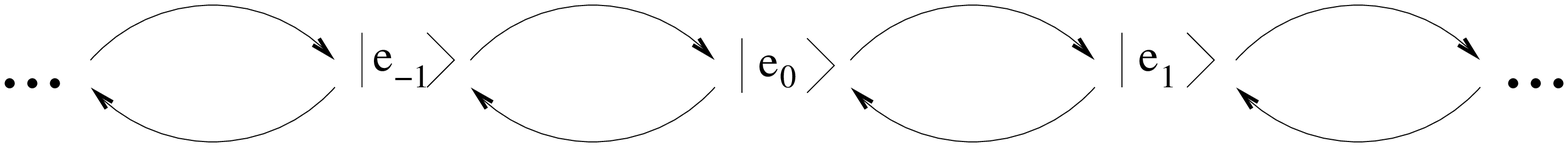}
\end{center}
With this directed graph, we can talk about quantum random walk over the Hilbert space  $\mathcal{H}$, the classical analogue
being the classical random walk over $\mathbb{Z}$, and embed $\mathcal{H}$ into a Markov $L$-coalgebra,  $\mathcal{H}_\mathbb{Z}$.
Now we introduce over the basis $\mathcal{H}_\mathbb{Z}$, the fiber $M_2(\mathbb{C})$ to
get the trivial tensor bundle $\mathcal{H}_\mathbb{Z} \otimes M_2(\mathbb{C})$.
We have said that the classical random walk over $\mathbb{Z}$ is isomorphic to the symbolic dynamic described by the map $x \mapsto 2x \mod 1$
over the interval $[0,1]$.
By fixing the Markov partition $(E_1, E_2)$ leading to the $(2,1)$-De Bruijn graph, we code
the periodic orbits of the classical dynamical system in an one-to-one correspondance with the periodic
orbits of this graph.
The strategy now consists to fix an unitary matrix $U$ and to consider
the operators $P$ and $Q$, such that $U=P+Q$, with:
\[ P = \frac{1}{\sqrt{2}}\begin{pmatrix}
 \alpha & \beta\\
0 & 0
\end{pmatrix}
\textrm{
and} \
Q= \frac{1}{\sqrt{2}}\begin{pmatrix}
 0 & 0\\
\gamma & \delta
\end{pmatrix}.
\]
We consider non commutative polynomials in $P$ and $Q$, i.e. the algebra $\mathbb{C}\bra P, \ Q \ket $
and denote by $\mathcal{D_{-}}, \ \mathcal{D_{+}}$ the dispersion operators, i.e. the linear maps
$$\mathcal{D_{-}}, \ \mathcal{D_{+}} : \ \mathcal{H}_\mathbb{Z} \otimes \mathbb{C}\bra P, \ Q \ket  \xrightarrow{} \mathcal{H}_\mathbb{Z} \otimes \mathbb{C}\bra P, \ Q \ket $$
which are defined for all $k \in \mathbb{Z}$ and for all discrete time $n \in \mathbb{Z}$,
$$\mathcal{D_{-}}( \  \vert e_{k+1} \ket \otimes  \Xi_{[k+1;n]} \ ) =  \vert e_{k} \ket \otimes  \Xi_{[k+1;n]}P
, \ \mathcal{D_{+}}( \  \vert e_{k-1} \ket \otimes  \Xi_{[k-1;n]} \ ) =  \vert e_{k} \ket \otimes  \Xi_{[k-1;n]}Q, $$
where $ \Xi_{[0;0]} = id$, $ \Xi_{[-1;1]} = P$, $ \Xi_{[+1;1]} = Q$ and so on.
The dynamics is defined by:
$$ \vert e_k \ket \otimes  \Xi_{[k;n+1]} :=  \mathcal{D_{-}}( \  \vert e_{k+1} \ket \otimes  \Xi_{[k+1;n]} \ )
+ \mathcal{D_{+}}( \  \vert e_{k-1} \ket \otimes  \Xi_{[k-1;n]} \ ). $$
That is:
\begin{center}
\includegraphics*[width=11cm]{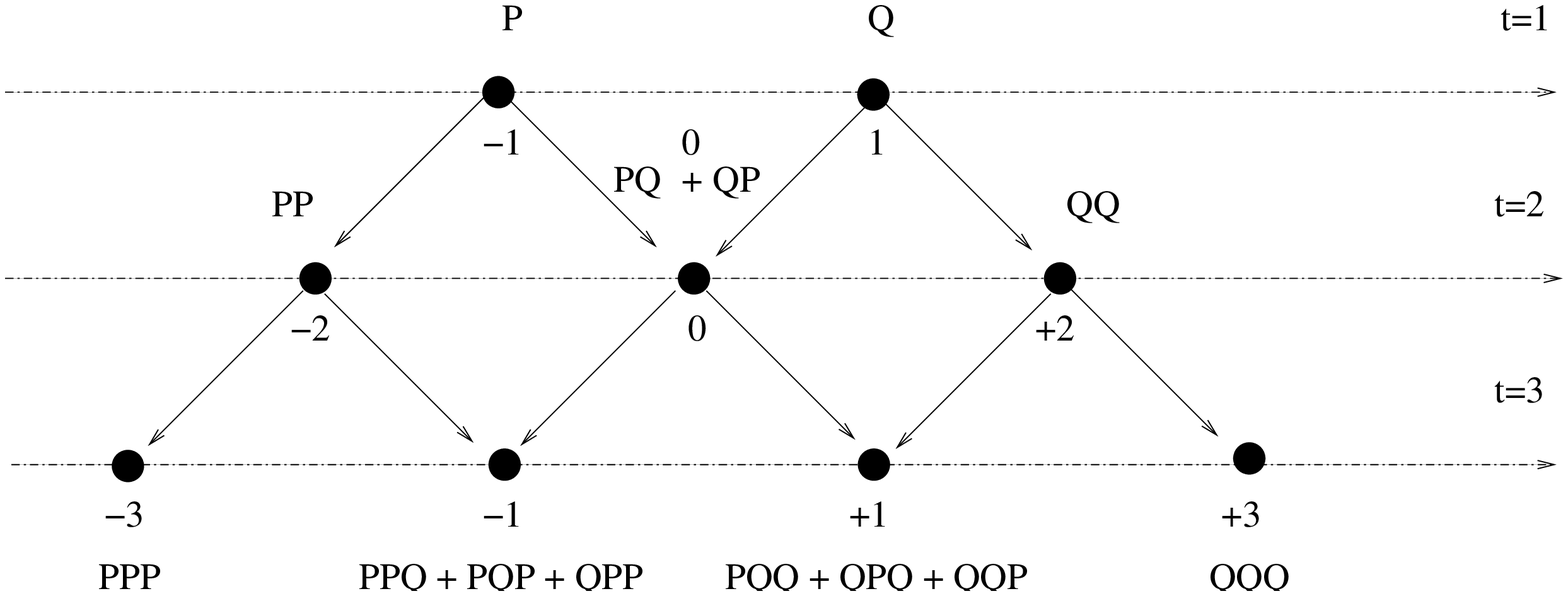}

\textsf{Quantum random walk over $\mathbb{Z}$, up to $t=3$.}
\end{center}
\begin{exam}{}
We yield here the non null polynomials $ \Xi_{[k;n+1]}$ for time $t = 0, \ldots, 4$. At time $t=0$, we get
$ \Xi_{[0;0]} = id$. At time $t=1$, $ \Xi_{[-1;1]} = P$, $ \Xi_{[+1;1]} = Q$. At time $t=2$, $ \Xi_{[-2;2]} = P^2$,
$ \Xi_{[0;2]} = PQ + QP$ and $ \Xi_{[+2;2]} = Q^2$. At time $t=3$, $ \Xi_{[-3;3]} = P^3$, $ \Xi_{[-1;3]} = QP^2 + PQP + P^2Q$,
$ \Xi_{[+1;3]} = PQ^2 + QPQ + Q^2P$ and $ \Xi_{[3;3]} = Q^3$. At time $t=4$,
$ \Xi_{[-4;4]} = P^4$, $ \Xi_{[-2;4]} = QP^3 + PQP^2 + P^2QP + P^3Q$, $ \Xi_{[0;4]} = P^2Q^2 + PQPQ + PQ^2P + Q^2P^2 + QPQP + QP^2Q$,
$ \Xi_{[+2;4]} = PQ^3 + QPQ^2 + Q^2PQ + Q^3P$ and $ \Xi_{[+4;4]} = Q^4$.
\end{exam}

Let us denote by $S(\mathbb{C}^2)$, the set of vectors $\psi$ of $S(\mathbb{C}^2)$ such that $\psi^\dagger \psi=1$.
The quantum random walk over a state $\psi \in S(\mathbb{C}^2)$ is defined by
the initial condition $ \Psi_{space = 0, time = 0} := \vert e_0 \ket \otimes \psi$. At time $n$, this state will spread and the probability
amplitude at position $k$ described by $\vert e_k \ket$ will be
$ \Psi_{k, n} :=  \vert e_k \ket \otimes  \Xi_{[k;n+1]}\psi$,  (since $P^\dagger P + Q^\dagger Q = I$, the norm of the initial state is preserved.).
We have an action from the bundle $\mathcal{H}_\mathbb{Z} \otimes \mathbb{C}\bra P, \ Q \ket$ on $\mathbb{C}^2$ described by
$$ \mathcal{RW}: \ \mathcal{H}_\mathbb{Z} \otimes \mathbb{C}\bra P, \ Q \ket \times  \mathbb{C}^2  \xrightarrow{} \ \mathcal{H}_\mathbb{Z} \otimes \mathbb{C}^2;
\ \ \ \ (\vert e_k \ket \otimes  \Xi_{[k;n+1]}, \psi)  \mapsto \vert e_k \ket \otimes  \Xi_{[k;n+1]}\psi.$$
The total state is $\Psi_{total}^n := \sum_k \Psi_{k, n}$.
\begin{prop}
For all $x \in \mathbb{C}\bra P, \ Q \ket $,
we define the right polynomial multiplication
$R_x: \mathbb{C}\bra P, \ Q \ket \xrightarrow{} \mathbb{C}\bra P, \ Q \ket $, $y \mapsto yx$, we have,
$[\mathcal{D_{+}}, \ \mathcal{D_{-}}] = id \otimes R_{[Q,P]}.$
\end{prop}
\Proof
Straightforward.
\eproof
\Rk \textbf{[On the importance of discrete quantum random walk]}

\noindent
Let us denote by $S(M_2(\mathbb{C}))$, the set of density matrices $\rho$ of $S(M_2(\mathbb{C}))$, i.e.
the set of trace one positive matrices.
We know that such a density matrix can be written by $\rho = \sum_i p_i \vert \psi_i \ket \bra \psi_i \vert$, where the $\psi_i \in S(\mathbb{C}^2)$
and $\sum_i p_i =1$.
We can apply the quantum random walk over each $\psi_i $, with different initial spacial conditions. By a suitable normalisation,
we recover a quantum channel which scatters each initial component of the initial density matrix $\rho$.
\section{Non commutative polynomials and the reading of periodic orbits of $G(\E)$ }
\Rk
From now on, we forget the algebraic relation between $P$, $Q$ and their powers. These monomials will be treaded simply
as non commutative words.

The aim of this section is to recover the polynomials $\Xi_{[k;n]}$, involved in the quantum random walk,
from the periodic orbits of the chaotic map $x \mapsto 2x \mod 1$. We have seen that the
classical random walk over $\mathbb{Z}$ is isomorphic to
the chaotic map $x \mapsto 2x \mod 1$, that all periodic orbits of this map were coded bijectively
into the $(2,1)$-De Bruijn graph,
associated with the unistochastic process $B_2$ and that
the extension of this graph leads to the directed graph associated with the coalgebra $\E$.
\begin{center}
\includegraphics*[width=4cm]{graph2.eps}
\end{center}
We will show that the periodic orbits
of this directed graph allow us to recover the combinatorics generated by the quantum random walk over $\mathbb{Z}$ and that this combinatorics
is governed by the coassociative coproduct of $\E$. We recall that this coproduct is defined by:
$$
\Delta a = a \otimes a + b \otimes c, \ \
\Delta b = a \otimes b + b \otimes d, \ \
\Delta c = d \otimes c + c \otimes a, \ \
\Delta d = d \otimes d + c \otimes b.
$$
However this directed graph can be also embeded into its natural Markov $L$-coalgebra. We define only the right coproduct $\Delta_M$.
We recall that by definition,
$$
\Delta_M a = a \otimes (a + b) \ \
\Delta_M b = b \otimes (c + d), \ \
\Delta_M c = c \otimes  (a +b), \ \
\Delta_M d = d \otimes (c + d).
$$
Let us start with the definition of a language.
\begin{defi}{[Language]}
We denote by $\F$ the free vector space generated by  all the words constructed from the alphabet $a,b,c,d$ and
representing a path of the graph $G(\E)$, for instance $aabdcabcb$. (It is also called the path space). We will call
such a space a {\it{language}}.
Its grammar, i.e the laws allowing us to construct words longer and longer
is constructed from the markovian coproduct of $G(\E)$, i.e. the substitution rules are:
$$a \mapsto ab, \ a \mapsto aa, \
b \mapsto bc, \ b \mapsto bd, \
c \mapsto ca, \ c \mapsto cb, \
d \mapsto dc, \ d \mapsto dd.$$
\end{defi}
\Rk
In (classical) theoretical computer science the definition of a grammar is more restrictive. Here, as in \cite{Malyshev},
we prefer view a grammar as a way to produce sequences of strings, words, via substitution rules. The language so obtained, is
then the space generated by such a grammar.
\Rk
As we associate with each tensor product $x \otimes y$ a directed arrow $x \xrightarrow{} y$, the relationship between the
$(2,1)$-De Bruijn graph, whose vertex set is $\{ P,Q \}$, and its extension is given by identifying
$ a := P \otimes P$, $b := P \otimes Q$, $c:= Q \otimes P$  and $d := Q \otimes Q $.

Often, for simplifying notation and only in the language $\F$, we will write $xy$ instead of $x \otimes y$.
No confusion is possible since, in the sequel we forgot the algebraic relations between letters.

We define now the contraction map. It will help us to go from the language $\F$ to the $(2,1)$-De Bruijn graph.
\begin{defi}{[Contraction map]}
For all $n>2$, the {\it{contraction map}} is the linear map
$$\mathcal{C}: \ \mathcal{F} \xrightarrow{} \mathbb{C}\bra P, \ Q \ket,
(y_1 \otimes y_2) \otimes (y_2 \otimes y_3) \otimes (y_3 \otimes y_4) \ldots (y_{n-1} \otimes y_n) \mapsto y_1y_2 \ldots y_n,$$ where
the $y_i$ stands for $P$ and $Q$. For $n =2$, the contraction is by convention equal to the usual product of $M_2( \mathbb{C} )$.
\end{defi}
\begin{exam}{}
For instance, the contraction of $ a \otimes b \otimes c:= (P \otimes P) \otimes (P \otimes Q) \otimes (Q \otimes P) $ is equal to $PPQP$.
\end{exam}
\Rk
The contraction is associative and generalize the usual product on $M_2( \mathbb{C} )$ denoted by $m$. In our example, it symbolizes
graphically that the gluing between the loop indexed by $P$ and the edge $P \xrightarrow{} Q$ at the vertex $P$ in the $(2,1)$-De Bruijn graph is
represented by the arrow $a \otimes b$ in $\mathcal{F}$.
\begin{prop}
For a time $t>1$, to any monomial $\Xi$ constructed from $P,Q$ in the algebra $\mathbb{C}\bra P, \ Q \ket $, excepted of course any
linear superposition of $P$ and $Q$,  corresponds an unique word $\omega$ in the language $\F$
such that $\mathcal{C}(\omega) = \Xi$.
\end{prop}
\Proof
Any sequence $\Xi$ constructed from $P,Q$ in the algebra $\mathbb{C}\bra P, \ Q \ket $ corresponds to an unique path of the $(2,1)$-De Bruijn graph, i.e. a
unique path of its extension.
\eproof

\noindent
Denote by
$(\Delta_M)_0 =id$, $(\Delta_M)_1 := \Delta_M $, $(\Delta_M)_2 := (id \otimes  \Delta_M) \Delta_M $,
more generally, for all $n>1$,
$(\Delta_M)_n := (\underbrace{id \otimes \ldots \otimes id}_{n-1} \otimes \Delta_M) (\Delta_M)_{n-1}, $
similarly for $\Delta$.

Let us show that the combinatorics generated by the quantum random walk
can be obtained by contraction of all the words
from the language $\F = \{ (\Delta_M)_n (a+b+c+d), n \in \mathbb{N} \}$ and that $\F$ can be also viewed as
equipped by the grammar \footnote{With these substitution laws, comming from a coassociative coproduct, the language $\F$
will be also said a {\it{coassociative language}}.} generated by $\Delta$, i.e.
$$
a  \mapsto  aa, a  \mapsto bc, \
b \mapsto ab, b  \mapsto bd, \
c \mapsto dc, c \mapsto ca, \
d \mapsto dd, d  \mapsto cb,
$$
i.e. $\F = \{ \Delta_n (a+b+c+d), n \in \mathbb{N} \}$. That is
we will show that all the words from the language $\F$, present at a fixed time $t=n >1$ and thus obtained by $(\Delta_M)_{n-2} \ (a+b+c+d)$,
can be also obtained by computing $\Delta_{n-2} \ (a+b+c+d)$.
\begin{lemm}
\label{z}
$\Delta_M (a+b) = \Delta(a+b) $ and $\Delta_M (c+d) = \Delta(c+d) $.
\end{lemm}
\Proof
Straightforward.
\eproof
\begin{lemm}
\label{zz}
If $x$ stands for $a,b,c$ or $d$,
we have the following equalities:
$$ \mathcal{C}( x \otimes a) P = \mathcal{C}( x \otimes a \otimes a ); \ \ \ \  \mathcal{C}( x \otimes a) Q = \mathcal{C}( x \otimes a \otimes b )$$
$$ \mathcal{C}( x \otimes b) P = \mathcal{C}( x \otimes b \otimes c ); \ \ \ \  \mathcal{C}( x \otimes b) Q = \mathcal{C}( x \otimes b \otimes d )$$
$$ \mathcal{C}( x \otimes c) P = \mathcal{C}( x \otimes c \otimes a ); \ \ \ \  \mathcal{C}( x \otimes c) Q = \mathcal{C}( x \otimes c \otimes b )$$
$$ \mathcal{C}( x \otimes d) P = \mathcal{C}( x \otimes d \otimes c ); \ \ \ \  \mathcal{C}( x \otimes d) Q = \mathcal{C}( x \otimes d \otimes d )$$
\end{lemm}
\Proof
Straightforward.
\eproof
\begin{coro}
Let $x$ stands for $a,b,c,d$. We have $\mathcal{C}(x) (P+Q) =\mathcal{C}(\Delta_M x)$.
\end{coro}
\Proof
Straightforward.
\eproof

These lemmas claim that for a fixed time $t=n > 1$ and a given vertex $k$ of $\mathbb{Z}$, all the polynomials present at this vertex can be either
recovered by those which were at the vertex $k-1$ and $k+1$ at time $t=n-1$ by multiplying by $P$ and $Q$ or by computing the
walk starting from $a,b,c,d$ and generated by the Markov coproduct $\Delta_M$, the expected polynomials
in $P,Q$ being obtained by contraction.
\begin{theo}
$(id \otimes \Delta) \Delta_M = (id \otimes \Delta_M) \Delta_M.$
\end{theo}
\Proof
Straightforward, thanks to the lemma \ref{z}.
For instance $a \xrightarrow{\Delta_M } a \otimes (a+b) \xrightarrow{id \otimes \Delta_M(a+b) }a \otimes  \Delta (a+b).$
\eproof
\begin{coro}
$(id \otimes \Delta) \Delta (a+b) = (id \otimes \Delta_M) \Delta_M(a+b),$ and
$(id \otimes \Delta) \Delta (c+d) = (id \otimes \Delta_M) \Delta_M(c+d).$
This implies that $ \forall n \ (\Delta_M)_n (a+b +c +d)= (\Delta)_n (a+b +c+d)$.
\end{coro}
\Proof
Straightforward, thanks to the lemma \ref{z}.
\eproof
\Rk
This corollary means that all the polynomials, in $P,Q$ created by the quantum random walk can be obtained by contraction of the
markovian walk over the directed graph $G(\E)$ or can be also obtained by contraction of the
coassociative walk over $G(\E)$.
Thus we have proved that the combinatorics generated by the quantum random walk can be viewed by a coassociative coproduct point of view.
We represent here the beginning of the walk coded in terms of the language $\F$.
\begin{center}
\includegraphics*[width=13cm]{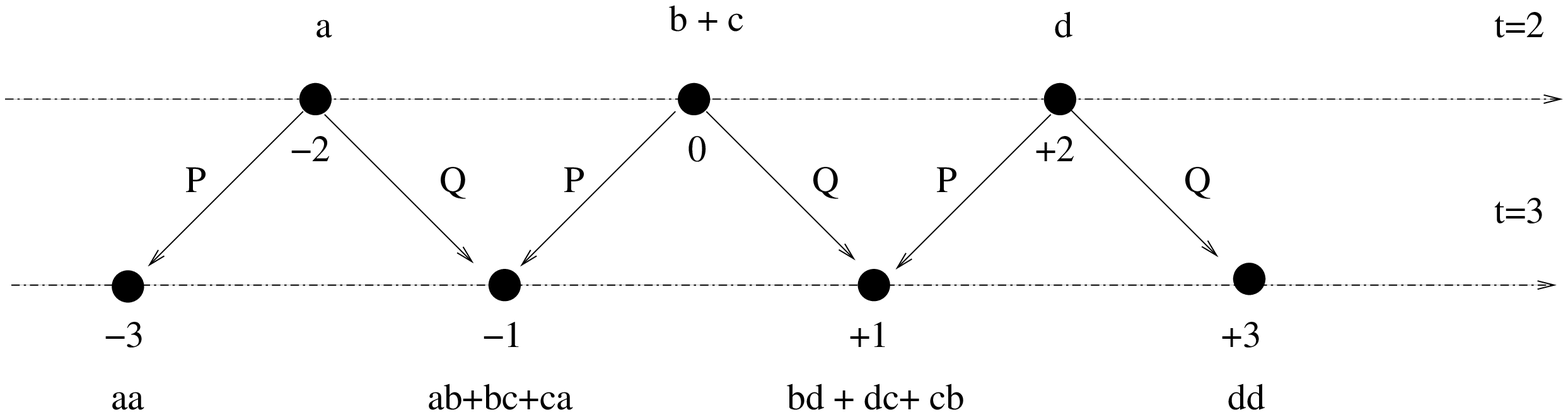}

\textsf{The quantum random walk, coded in terms of the language $\F$.}
\end{center}
For the moment, we get all the words created by the coassociative language. But if a word is picked up from this language how can we say
that it has to belong to such or such vertex?
We have to enlarge the definition of the language $\F$ by defining an index map and an index language.
From now on, we denote, by convention, $x_{-1,-1} :=a$, $x_{-1,+1} :=b$, $x_{+1,-1} :=c$, $x_{+1,+1} :=d$. Notice then that
a word from the language $\F$ can be written like $\omega := x_{i_1i_2}x_{i_2i_3} \ldots x_{i_{n-1}i_n}$.
The index language $\hat{\F}$ is by definition $ \mathbb{Z} \otimes \F$.
\begin{defi}{[Index map]}
Let $\omega$ be a word from the language $\F$, i.e, $\omega := x_{i_1i_2}x_{i_2i_3} \ldots x_{i_{n-1}i_n}$.
we define the (linear) {\it{index map}} as:
$\widehat{\ind}: \F \xrightarrow{}\hat{\F} $, $\omega \mapsto (\ind(\omega) \otimes \omega)$,
with $ \ind (\omega) = \ind (x_{i_1i_2}x_{i_2i_3} \ldots x_{i_{n-1}i_n}) := \sum_{k=1} ^n \ i_k$.
\end{defi}
\begin{prop}
Let $\omega := x_{i_1i_2} \ldots x_{i_{n-1}i_n}$ be a word from the language $\F$. The index
$\ind(\omega)$ is equal to the number of $Q$ minus the number of $P$ obtained in the contraction of the word $\omega$. Therefore,
the index map fixes the vertex attributed by the quantum random walk over $\mathbb{Z}$.
\end{prop}
\Proof
We will proceed by recurrence. It is true for $n=2$, i.e. for $a,b,c,d$. Suppose $\omega$, a word present at vertex $k$ and at time $t=n > 2$. We have
$\omega  = x_{i_1i_2} \ldots x_{i_{n-1}i_n}$, and the index $\ind(\omega ) =k$ does indicate
the number of $Q$ minus the number of $P$ obtained in the contraction of this word.
At time $t=n+1$, $\omega \mapsto \omega \otimes x_{i_n,i_{n+1}}$. By definition of the quantum random walk this word will
be at vertex $k+1$ if $x_{i_n,i_{n+1}}$ is equal to $Q$, or $k-1$ if $x_{i_n,i_{n+1}}$ is equal to $P$. Now
$\ind(\omega \otimes x_{i_n,i_{n+1}})=\ind(\omega) + i_{n+1}$. By definition, $i_{n+1}= +1$ for $b$ and $d$, which
are words finishing by $Q$ and $i_{n+1}= -1$ for $a$ and $c$ which
are words finishing by $P$.
\eproof
\begin{exam}{}
$ind(a) = -2$ and $\mathcal{C}(a) = P^2$. Hence the word $a$ has to be present at time $t=2$. Moreover
its contraction yielding the monomial $P^2$, $a$ is at vertex $-2$, as expected.
\end{exam}

The next question is how can we produce all these words, coming from the language $\F$, from the notion of
periodic orbits of the classical chaotic map $x \mapsto 2x \mod 1$.

\begin{defi}{[Periodic orbits, pattern]}
We define an equivalence relation $\sim$ inside $\F$ by saying that
$\omega_1 \sim \omega_2$ iff $\omega_1 =  x_{i_1i_2}x_{i_2i_3} \ldots x_{i_{n-1}i_1}$, for some $n$
and $\exists m, \ \tau^m( \omega_1) =\omega_2$.
The set $\PO = \F / \sim $ is the set of the {\it{periodic orbits}} of the directed graph $G(\E)$. We denote by $< \omega>$ the {\it{pattern}}
of an equivalence
classe associated with $\omega$ and its permutations, i.e. $< \omega > :=  < x_{i_1i_2}x_{i_2i_3} \ldots x_{i_{n-1}i_1}>$.
A {\it{periodic orbit}} is just the graphical representation of the pattern. Often, we will confound the two words.
\Rk
Fix a time $t$. We will see later that the lenght of the pattern of a periodic orbit $<\omega>$ present at $t$ denoted by $l(<\omega>)$, will
be equal to $t$.
\end{defi}
\begin{exam}{}
We have
$a \otimes b \otimes c \ \sim \ c \otimes a \otimes b \ \sim \ b \otimes c \otimes a$. The equivalent classe
is designed by the pattern $< a \otimes b \otimes c >$ and the associated periodic orbit is $\ldots abcabcabcabc\ldots$.
\begin{center}
\includegraphics*[width=1.8cm]{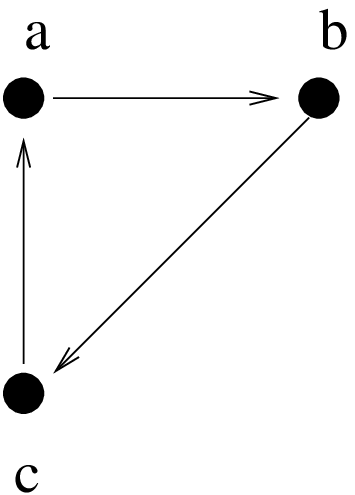}

\textsf{Periodic orbit, $< a \otimes b \otimes c >$, with pattern of lenght 3.}
\end{center}
This very periodic orbit can be also represented by the pattern $< a \otimes b \otimes c \otimes a \otimes b \otimes c >$, i.e we cover
two times the triangle and that is all.
\end{exam}
Similarly with the language $\F$, we have to enlarge the vector space of the periodic orbits $\PO$ to keep the notion of vertex
attributed by the quantum walk to each periodic orbit. We denote by $\widehat{\PO} := \mathbb{Z} \otimes \PO$ such a set.
\begin{defi}{[Index map in $\PO$]}
We define the (linear) {\it{index map}} $\widehat{\Ind}: \ \PO \xrightarrow{} \widehat{\PO}$, $< \omega > \mapsto \Ind(< \omega >) \otimes < \omega >$
with,
$$ \Ind(< \omega >) :=  \Ind (< x_{i_1i_2}x_{i_2i_3} \ldots x_{i_{n-1}i_1}>) := \frac{1}{2} ((i_1 + i_2) + (i_2+ i_3) + \ldots  (i_{n-1} + i_1))
= \sum_{k=1}^{n-1}i_k .$$
\end{defi}
\begin{exam}{}
$\Ind(x_{-1,-1}) = -1$, $\Ind(x_{-1,+1}) = 0$, $\Ind(x_{+1,-1}) = 0$, $\Ind(x_{+1,+1}) = +1$.
\end{exam}
This definition does not depend on the choice of the representative of the equivalent classe.
Once we have the definition of periodic orbits, we have to read them to obtain information.
\begin{defi}{[Reading map]}
Let $< \omega> := < x_{i_1i_2}x_{i_2i_3} \ldots x_{i_{n}i_1}>$ be a periodic orbit.
The {\it{reading map}} is denoted $R: \PO \xrightarrow{} F$ with $x_{i_1i_2}x_{i_2i_3} \ldots x_{i_{n}i_1} \mapsto
\sum_{k=1}^{n} x_{i_k i_{k+1}}x_{i_{k+1}i_{k+2}} \ldots x_{i_{n+k-2}i_{k+n -1}}$, the labels being understood modulo $n$.
The reading map does not depend on the choice of the representative of the equivalent classe.
\end{defi}
\Rk
As we are interested in the notion of information generated by the reading of periodic orbits, we wish to avoid the repetition
of words, because knowing that some words are present several times in the reading of a periodic orbit does not bring
more information. That is why we define the map $J : \ \F \xrightarrow{} \F$, $\sum_k \lambda_k \omega_k \mapsto \sum_k \omega_k$,
where the $\lambda_k$ are scalars (integers).
From now on, the (linear) reading map $R$ will be always composed with the (non-linear) map $J$. This composition will be still denoted by $R$.
\begin{exam}{}
Consider the periodic orbit $<abc>$. Its index is $-1$ and its reading yields
$ab + bc +ca$. By contraction we obtain $PPQ + PQP + QPP$, which is exactely the polynomial expected at time
$t=3$ and at vertex $-1$.
\end{exam}
\begin{exam}{}
\label{lk}
The reading of the periodic orbit $\bra b \otimes c \otimes b \otimes c \ket$, with a pattern of lenght 4, yields
$b\otimes c\otimes b + c\otimes b\otimes c + b\otimes c\otimes b + c\otimes b\otimes c$. Composed by the map $J$ we get
$R \bra b \otimes c \otimes b \otimes c \ket := b\otimes c\otimes b + c\otimes b\otimes c$. By contraction, we obtain $PQPQ +QPQP$.
Moreover its index is $0$.
\begin{center}
\includegraphics*[width=1.8cm]{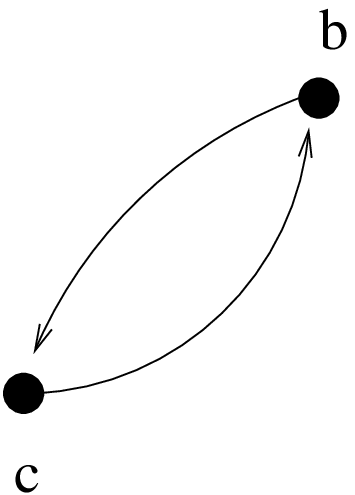}

\textsf{Periodic orbits of period 4, $<bcbc>$, with pattern of lenght 4.}
\end{center}
The reading of the periodic orbit $\bra a \otimes b \otimes d \otimes c \ket$, with pattern of
lenght 4, yields $a \otimes b \otimes d +  b \otimes d \otimes c +  d \otimes c \otimes a + c \otimes a \otimes b $.
By contraction, we obtain $PPQQ + PQQP + QQPP + QPPQ$. Its index is $0$.
\begin{center}
\includegraphics*[width=1.8cm]{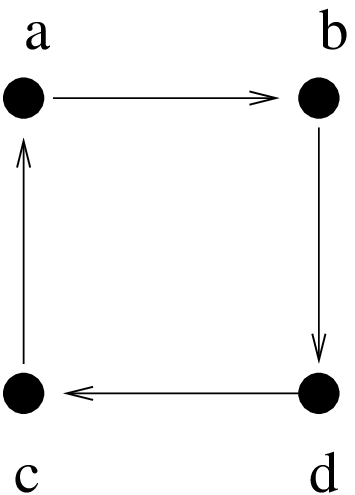}

\textsf{Periodic orbit, $<abdc>$, with pattern of lenght period 4.}
\end{center}
That is the sum of these two periodic orbits yield the
polynomials expected at time
$t=4$ and at vertex $0$.
\end{exam}
So as to give a minorant of the number of periodic orbits at a given time $n$ at vertex $k$, we have to link the number $k$ to the
number $x_P$ of $P$ and the number $x_Q$ of $Q$. Suppose $k$ positive \footnote{As the walk is symetric, we have as many
polynomials at the vertex $k$ as at the vertex $-k$.}.
This means that $x_Q = x_P + k$. As $x_P + x_Q = n$, we get $x_P = \frac{n -k}{2}$. As this solution has to be an integer, this will
fix the possibilites of $k$, i.e. the possible vertices reached by the quantum random walk at time $n$. We denote $\kappa : =\frac{n -k}{2}$.
\begin{prop}
Set $\varpi := \frac{1}{n}\frac{n!}{\kappa  !(n- \kappa )!}$ .
The number of periodic orbits at the vertex $k$ and at time $n$ is greater or equal to $\varpi$ if
this number is an integer and greater or equal to the integer part of $\varpi+ 1$ if $\varpi$
is not an integer.
\end{prop}
\Proof
For a given vertex $k$ and a given time $n$, we have to have $\frac{n!}{\kappa !(n- \kappa  )!}$ polynomials in $P$ and $Q$ and by definition
we know that the reading of these periodic orbits yields at most $n$ different words. \eproof

\noindent
In the following picture, we indicate the classical periodic orbits involved into the quantum walk over $\mathbb{Z}$.
\begin{center}
\includegraphics*[width=16cm]{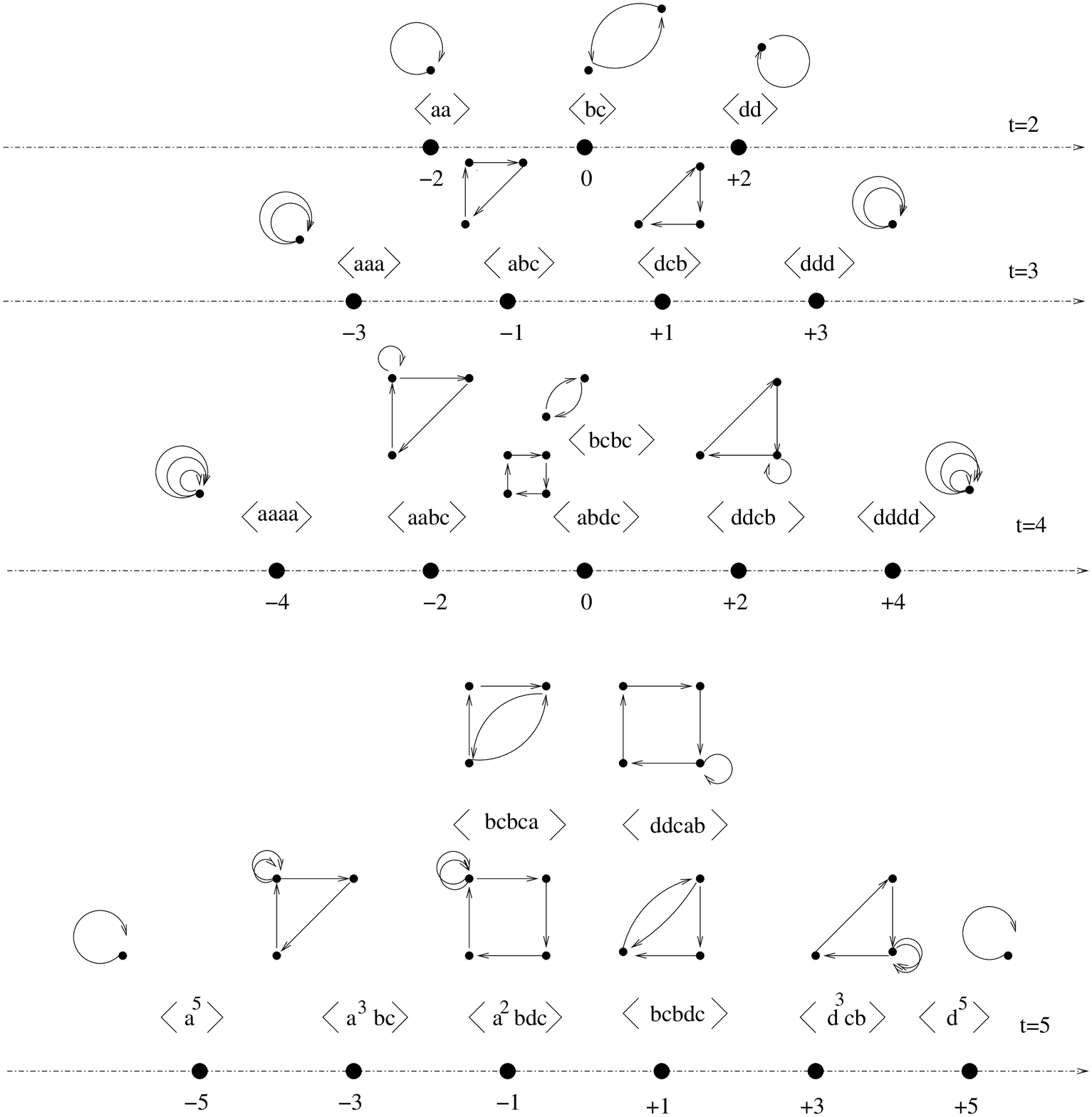}
\textsf{Periodic orbits and their associated pictures.}
\end{center}
\begin{prop}
Let $< \omega> := < x_{i_1i_2}x_{i_2i_3} \ldots x_{i_{n}i_1}>$ be the pattern of a periodic orbit.
Its reading yields $R< \omega>  := \sum_k ^n X_k$, where $X_k$ is
the word $x_{i_k i_{k+1}}x_{i_{k+1}i_{k+2}} \ldots x_{i_{k-2 + n \mod n}i_{k-1 + n \mod n}}$.
We have $\Ind( < \omega> ) :=  \ind(X_k)$, for all $k = 1, \ldots,n$
which proves that the reading map decomposes the periodic orbit into words with same index.
\end{prop}
\Proof
Let $< \omega> := < x_{i_1i_2}x_{i_2i_3} \ldots x_{i_{n}i_1}>$ be a periodic orbit and the $X_k$ its decomposition under the reading map.
We have $s_1 := \ind (X_1) := \ind x _{i_1i_2}x_{i_2i_3} \ldots x_{i_{n-1}i_n} = \sum_{k=1} ^n \ i_k$.
$s_2 := \ind (X_2) := \ind x _{i_2i_3}x_{i_3i_4} \ldots x_{i_{n-1}i_n}x_{i_{n}i_1} = \sum_{k=2} ^n \ i_k + i_1$. Thus $s_1 = s_2$,
$s_k:= \ind (X_k) := i_k  + i_{k+1} + i_{k+2} +  \ldots i_n + i_1 + \ldots i_{k-1} = s_1.$
Besides, by definition $\Ind(<\omega>) := \frac{1}{2}(2 \ind(\omega))$, whatever the choice of the representative is.
\eproof
\Rk
A periodic
orbit with index $k$, will be affected to the vertex $k$. Its reading will yield words of index $k$.
\begin{exam}{}
The reading of the periodic orbit $<abc>$ is: $R(<abc>) := ab + bc +ca$. We have $\Ind(<abc>) =-1$ and $\ind(ab) =\ind(ca)= \ind(bc) =-1.$
\end{exam}
\begin{defi}{[Completion]}
For a time $t=n$ and a given vertex $k$, all the polynomials can be recovered by the contraction of words coming from the
coassociative language $\F$ and having the same index $k$. Suppose we pick up one of the word present at this vertex $k$,
say $x_{i_1i_2}x_{i_2i_3} \ldots x_{i_{n-1}i_n}$. The {\it{completion}} $\Comp$ maps $x_{i_1i_2}x_{i_2i_3} \ldots x_{i_{n-1}i_n}$ into
$< x_{i_1i_2}x_{i_2i_3} \ldots x_{i_{n-1}i_n} x_{i_{n}i_{1}} >$, thus it is a map from $\F$ to $\PO$.
\end{defi}
\begin{prop}
$\Ind(\Comp (x_{i_1i_2}x_{i_2i_3} \ldots x_{i_{n-1}i_n}) )=  \ind(x_{i_1i_2}x_{i_2i_3} \ldots x_{i_{n-1}i_n}).$
\end{prop}
\Proof
Straightforward.
\eproof
\Rk
The reading of the completion of a word present at vertex $k$
will yield many words present at this vertex. By contraction we will recover polynomials
in $P,Q$ present at vertex $k$. As all these polynomials are coded bijectively by a word of $\F$ present at the
vertex $k$, the completion of all these words will form all the necessary periodic orbits whose the reading will yield all the words
present at $k$.
For the time being, we started from the polynomial algebra $\mathbb{C}\bra P, \ Q \ket $
and arrived at the periodic orbits set of the chaotic map $x \mapsto 2x \mod 1$.
We showed that the reading of all the patterns of periodic orbits present at a vertex $k$ yielded all the words of the language
$\F$ present at $k$ and whose the contraction gave the polynomials in $P,Q$.

The following idea is to equip the periodic orbits with the coproduct of $\E$. Thanks to this coproduct, we will able to speak of the growth
of periodic orbits and to recover the quantum random walk over $\mathbb{Z}$ from the reading of them.
\Rk
To avoid redundancy, we define the map $J_*: \PO \xrightarrow{} \PO$, $\sum_k \lambda_k <\omega_k> \mapsto \sum_k <\omega_k> $,
where $\lambda_k $ are scalars (integers).
\begin{defi}{[Growth of periodic orbit]}
We denote by $\mathcal{G}: \ \PO \xrightarrow{} \PO$ the {\it{growth operator}}:
$$\mathcal{G}(< \omega >) = \sum_{k=1}^{l(<\omega >)} \ <  \ ^k \delta(\omega) \ > ,$$
where $^k \delta = id \otimes \ldots \otimes \underbrace{\Delta}_k \otimes \ldots \otimes id$ and $l(<\omega >)$ is the lenght
of the pattern of the periodic orbit $<\omega >$.
\end{defi}
\Rk
As $\Delta x_{ij} := \sum_{k=+1,-1} \ x_{ik}\otimes x_{kj}$, it is easy to see that $\mathcal{G}: \ \PO \xrightarrow{} \PO$.
We take into account all the possible substitutions coming from the coassociative grammar.
\Rk
Notice also that the coproduct leaves a letter of index $k$ into two letters of index $k+1$ and $k-1$, i.e
the coassociative coproduct let the index invariant. For instance
$\Delta b = ab + bd$ and $\Ind(b) = 0 \mapsto (\Ind(ab)=-1) + (\Ind(bd)=+1)$.
\begin{theo}
\label{thn}
The growth operator applied on all the periodic orbits at time $t=n$ will yield all the periodic orbits at time $t=n+1$.
By applying the operator $J_*$, we will recover exactly the number of periodic orbits present at time $t=n+1$.
\end{theo}
\Proof
Let $\omega =x_{i_1i_2}x_{i_2i_3} \ldots x_{i_{n-1}i_n}$ a word present at time $t=n$, its completion yields
the periodic orbit
$<x_{i_1i_2}x_{i_2i_3} \ldots x_{i_{n-1}i_n}x_{i_{n}i_1}>$ whose reading
will give us the words $x_{i_k i_{k+1}}x_{i_{k+1}i_{k+2}} \ldots x_{i_{k+n-2 \mod n}i_{k+n-1 \mod n}}$, $k=1, \ldots, n$. By definition of
the quantum random walk, we have to multiply them by $P$ and $Q$ to have the new polynomials present at time $t=n+1$.
Let see how it works on $\omega$ itself. We get
$\mathcal{C}(x_{i_1i_2}x_{i_2i_3} \ldots x_{i_{n-1}i_n})P$ and $\mathcal{C}(x_{i_1i_2}x_{i_2i_3} \ldots x_{i_{n-1}i_n})Q$.
These two polynomials come from the contraction of two words present at time $t=n+1$,
$x_{i_1i_2}x_{i_2i_3} \ldots x_{i_{n-1}i_n} x_{i_{n}i_{m}}$ and $x_{i_1i_2}x_{i_2i_3} \ldots x_{i_{n-1}i_n} x_{i_{n}i_{m'}}$
with $m \not= m'$. By completion of these two words we get
$<x_{i_1i_2}x_{i_2i_3} \ldots x_{i_{n-1}i_n} x_{i_{n}i_{k}}x_{i_{m}i_{1}}>$ and
$<x_{i_1i_2}x_{i_2i_3} \ldots x_{i_{n-1}i_n} x_{i_{n}i_{k'}}x_{i_{m'}i_{1}}>$.
The sum of these two periodic orbits is obviously equal to $<x_{i_1i_2}x_{i_2i_3} \ldots  x_{i_{n-1}i_n} \Delta x_{i_{n}i_1}>$.
Now, for some $k$
and by reading the labels $\mod n$, the contraction of the word $x_{i_k i_{k+1}}x_{i_{k+1}i_{k+2}} \ldots x_{i_{k+n-2}i_{k+n-1}}$,
by multiplication by $P$ and $Q$ will come from the contraction of the word
$x_{i_k i_{k+1}}x_{i_{k+1}i_{k+2}} \ldots x_{i_{k+n-2}i_{k+n-1}}x_{i_{k+n-1}i_{m}}$
and the word  $x_{i_k i_{k+1}}x_{i_{k+1}i_{k+2}} \ldots x_{i_{k+n-2}i_{k+n-1}}x_{i_{k+n-1}i_{m'}}$.
The completion of these two words comes from
the periodic orbit
$<x_{i_k i_{k+1}}x_{i_{k+1}i_{k+2}} \ldots x_{i_{k+n-2}i_{k+n-1}}\\ \Delta x_{i_{k+n-1}i_{k}}>$.
The last term $\Delta x_{i_{k+n-1}i_{k }}$ is equal to $\Delta x_{i_{k-1}i_{k}}$. That is why
the growth operator works on periodic orbits to recover the words at time $t=n+1$ from the periodic orbits at time $t=n$.
We have just obtained all the periodic orbits with repetition, by applying  $J_*$, we get the theorem.
\eproof
\Rk
It is worth to notice that the markovian coproduct is closely related to the coassociative coproduct.
Indeed for  a word $x_{i_1i_2}x_{i_2i_3} \ldots x_{i_{n-1}i_n}$, the product by $P$ and $Q$ will yield
the words $x_{i_1i_2}x_{i_2i_3} \ldots \\ \Delta_M(x_{i_{n-1}i_n})$, see lemma \ref{zz}, whereas its completion will use the coassociative coproduct.
\begin{exam}{}
Consider the word $ab$. The contraction of $ab$ multiplied by $P$ and $Q$ yields $abc$ and $abd$.
Its completion will yield $abca$ and $abdc$, which is $<ab \Delta c>$.
Thus $ab \xrightarrow{P,Q} abc + abd = a \Delta_M(b) \xrightarrow{\Comp} \\ <ab \Delta c>$.
\end{exam}
\Rk
We have proved that all the polynomials of the quantum random walk over $\mathbb{Z}$ can be obtained
by reading periodic orbits of the classical chaotic map $x \mapsto 2x \mod 1$, and that $\PO$ can be viewed as
a coassociative language, since its substitution rules come from a coassociative coproduct.
\begin{exam}{[Reconstruction of the quantum walk from classical periodic orbits]}
We can start by the two loops $<aa>$ and $<dd>$, at $t=2$, because the patterns created by $<bc>$, with the growth operator can be recovered by these two loops.
At time $t=3$, the growth operators yield $<aa> \mapsto <aaa> + <bca> + <aaa>  + <abc>$. To avoid the redundancy of information, i.e. by applying
$J_*$, we consider
the periodic orbits $<aaa> + <abc>$. By applying the linear map $\widehat{Ind}$, we find that, $Ind(<aaa>)=-3$ and $Ind(<abc>)=-1$.
Their reading yields $aa + ab + bc +ca$ which are separated by their indexes. By applying the linear map $\widehat{ind}$, we find that
$aa$ has an index equal to $-3$ and $ab$, $bc$ and $ca$ are all of index $-1$. Similarly
$<dd> \mapsto <ddd> + <cbd> + <ddd>  + <cbd>$, therefore we consider the orbit $<ddd> + <cbd>$. Notice that only these 4 periodic orbits
are present at time $t=3$. By applying the growth operator at time $t=3$, we will still obtain all the orbits present at time $t=4$, and so forth.
\end{exam}
\subsection{Arithmetics from the periodic orbits of the map $x \mapsto 2x \mod 1$}
The aim of this section is to show that all these periodic orbits are generated
by 6 fondamental orbits, which are:
\begin{center}
\includegraphics*[width=10cm]{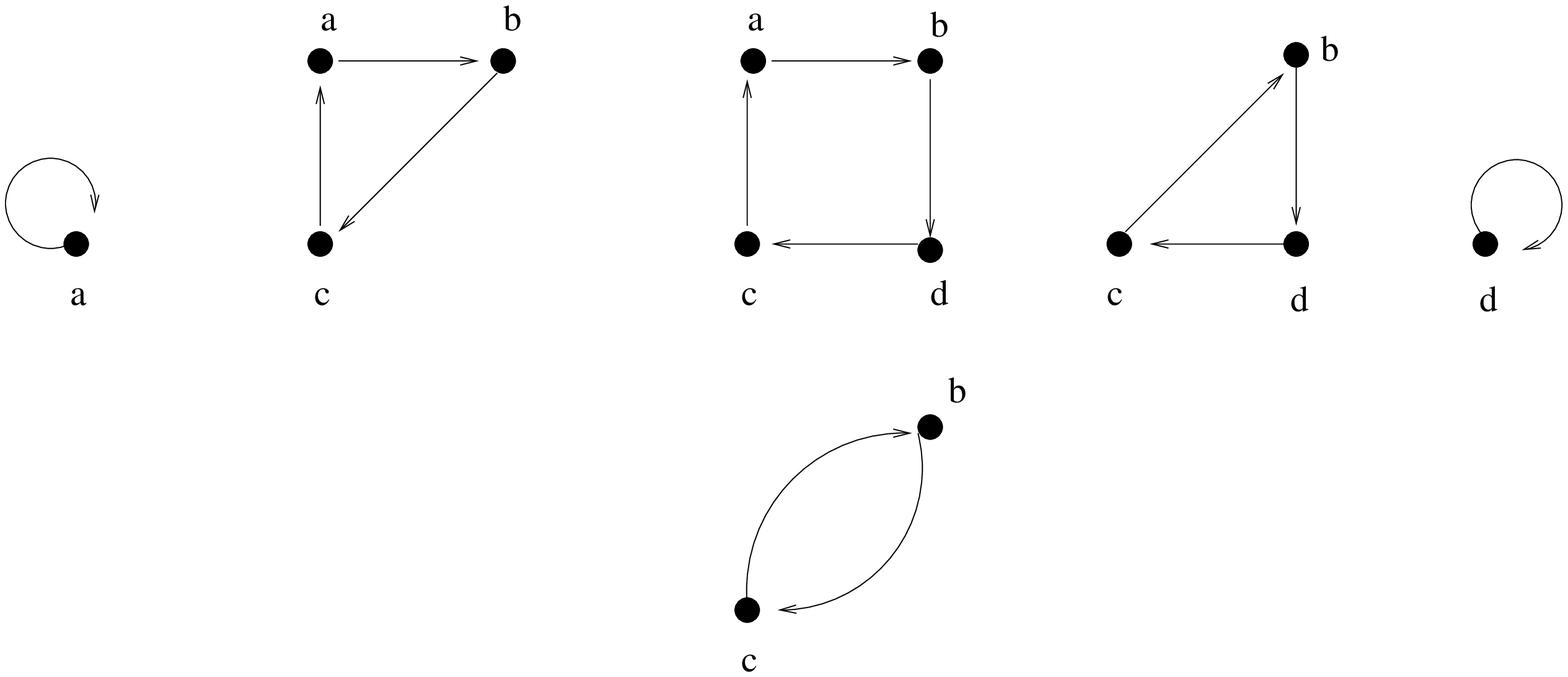}

\textsf{Fondamental periodic orbits and their associated pictures.}
\end{center}
They will play the same r\^ole that the prime numbers in the integer arithmetics. For that, we need a product.
\begin{defi}{}
Define the {\it{gluing map}} $\# \ : \PO \xrightarrow{} \PO $ which glues two periodic orbits with their graphic intersection.
\end{defi}
\begin{exam}{}
We show here how to glue periodic orbits.
\begin{center}
\includegraphics*[width=7cm]{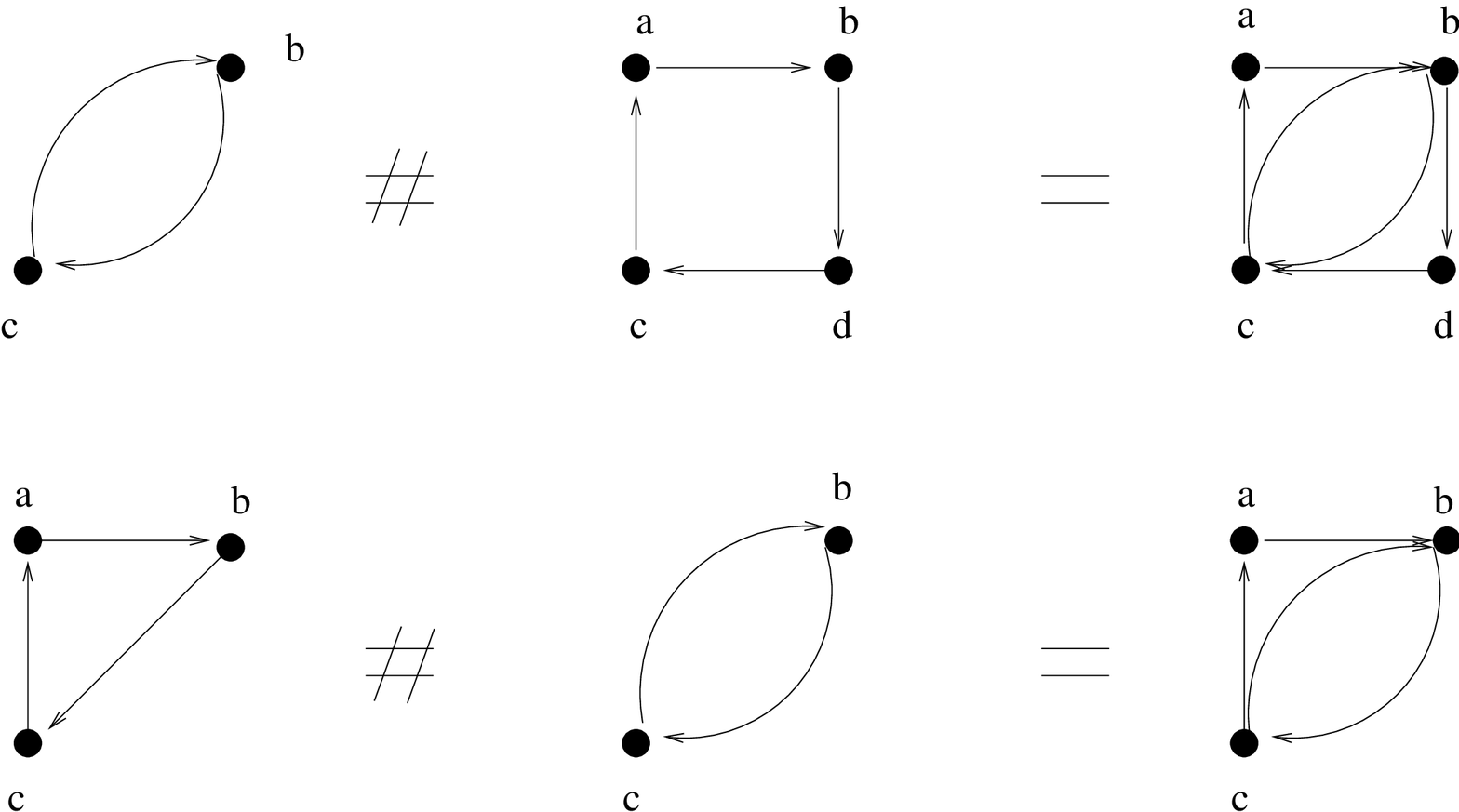}

\textsf{Example of  gluing of periodic orbits.}
\end{center}
\end{exam}
\begin{prop}
Any periodic orbits of $\PO$ can be decomposed into fondamental periodic orbits, i.e. the graphical representation
of a periodic orbit can be viewed as the gluing of some of these fondamental orbits.
\end{prop}
\Proof
All the periodic orbits are created by the growth operator from the two loops $<aa>$ and $<dd>$. Therefore,
it suffices to prove that the growth of any fondamental periodic orbits still yield fondamental orbits.
But $<aa>$ yields $<aaa>$ and $<abc>$,
$<dd>$ yields $<ddd>$ and $<cbd>$,
$<abc>$ yields $<aabc>$, $<bcbc>$ and $<abdc>$
$<cbd>$ yields $<ddcb>$, $<cbcb>$ and $<abdc>$
$<bcbc>$ yields $<abcbc>$, $<bdcbc>$
$<abdc>$ yields $<aabdc>$, $<bcbdc>$, $<abddc>$, $<abcbc>$.
All these periodic orbits are graphically the gluing of fondamentals orbits. For instance $<abddc>$ is the gluing of
the square $<abdc>$ and the loop $<dd>$, i.e. $<abddc> =<abdc> \# <dd>$.
$<abcbc>$ is the gluing of the triangle $<abc>$ with $<bc>$, i.e. $<abcbc> =<abc> \# <bc>$ and so on.
\eproof

\section{Conclusion}
To quantify the Bernoulli walk over $\mathbb{Z}$ we choose the following strategy. Firstly we use the isomorphism between
this discrete process and the chaotic map $x \mapsto 2x \mod 1$ with $x \in [0,1]$.
With this chaotic process, an unistochastic matrix $B_2$ has been associated \cite{Kus}. This matrix is identified with
the $(2,1)$-De Bruijn directed graph
which can be viewed as a Markov $L$-coalgebra.
This allows to code periodic orbits of this chaotic map in one to one way with the periodic orbits of the $(2,1)$-De Bruijn directed graph.
To quantify this system we choose an unitary matrix $U$ such that
the point by point product or Hadamard product between $U$ and $U\dagger$ yields $B_2$. This leads to the notion of quantum graph.
From a choice of matrix $U=P+Q$ among possible quantisations, we recover a non commutative algebra $\mathbb{C} \bra P, \ Q \ket$,
subalgebra of $M_2(\mathbb{C})$, the decomposition
of the unitary matrix yielding the matrices $P$ and $Q$ which generate the quantum random walk over $\mathbb{Z}$ \cite{Konno}.
But to study quantum chaos, we learn from \cite{Gutzwiller} that the r\^ole of the peridic orbits of the associated classical system
are capital. Would it be possible to yield a physical interpretation of the notion of coassociative
coproduct in the quantisation of a classical dynamical system?
We know that a classical dynamical system can be modelized by a commutative $C^*$-algebra.
To quantify such a system we look for a non commutative algebra embeding it.
However, in some quantum chaos experiments \cite{Gutzwiller}, the periodic orbits of the classical system
play a capital r\^ole. In fact, periodic orbits seem to be invariant when we quantify it.
As we show in the model of the quantum random walk over $\mathbb{Z}$, these classical periodic orbits (of the chaotic map $x \mapsto 2x \mod 1$)
are always present, not in the form of a classical trajectory, but in the form of a coassociative language. A physical interpretation
of the notion of coassociative coproduct would be to generate the grammar of the language associated with the periodic orbits. In addition to
the required non commutative algebra in the quantisation of a chaotic system, a coassociative coalgebra, closely related
to a Markov $L$-coalgebra would be required to manipulate
classical periodic orbits, via a coassociative grammar. To recover the combinatorics involved by the quantum dynamics
from the language of periodic orbits, we could use a kind of reading map and contraction map. The link between the combinatorics
generated by the quantum system and
and the coassociative language of its periodic orbits could be interpreted in terms of a smash biproduct \cite{Majid}\cite{Caenepeel}.

\noindent
\textbf{Acknowledgments:}
The author wishes to thank Dimitri Petritis for useful discussions and fruitful advice for the
redaction of this paper and to S. Severini for pointing him the precise definition of the De Bruijn graphs.\\

\bibliographystyle{plain}
\bibliography{These}

\end{document}